\pdfoutput=1 % for arXiv
\documentclass[aps,pra,superscriptaddress,twocolumn
]{revtex4-2}
\usepackage[]{amsmath,amssymb,mathrsfs,titlesec,physics,amsfonts}

\usepackage{ascmac,bm}
\usepackage[pdftex]{graphicx}

\usepackage{natbib}
\usepackage{url,textcase}

\usepackage[svgnames%,psnames %for arXiv
]{xcolor}
\usepackage[colorlinks,citecolor=Green,linkcolor=Blue,linktocpage,unicode]{hyperref}

  \usepackage{amsthm,dsfont}

\newcommand{\e}{\mathrm{e}}
\renewcommand{\i}{\mathrm{i}}

\renewcommand{\d}{{\rm d}}

\newcommand{\majrev}[1]{{#1}}

\begin{document}
\title{Quantum feedback cooling of a trapped nanoparticle by using a low-pass filter}
\author{Shuma Sugiura}
 \email{sugiura@cat.phys.s.u-tokyo.ac.jp}
 \affiliation{%
Department of Physics, University of Tokyo, 7-3-1 Hongo, Bunkyo-ku, Tokyo, 113-8654, Japan}
\author{Masahito Ueda}%
\affiliation{%
Department of Physics, University of Tokyo, 7-3-1 Hongo, Bunkyo-ku, Tokyo, 113-8654, Japan}
\affiliation{
 Institute for Physics of Intelligence, University of Tokyo, 7-3-1 Hongo, Bunkyo-ku, Tokyo, 113-0033, Japan}
 \affiliation{
 RIKEN Center for Emergent Matter Science (CEMS), Wako, Saitama, 351-0198, Japan
}%
\date{\today}

\begin{abstract}
  We propose a low-pass-filter (LPF) feedback control for cooling a trapped particle with a low-pass filter, 
  \majrev{which utilizes a shift of the potential caused by the feedback operation.
 By incorporating this shift in the {energy cost function}, 
 we} show that the LPF control can achieve the minimum phonon occupation number that is lower 
 than {cold damping with a band-pass filter, that with delayed feedback, 
 and linear--quadratic--Gaussian (LQG) control, 
 the last two of which} are the standard methods of ground-state cooling of a levitated nanoparticle. 
 For the detection efficiency of $90\%$, the achievable phonon occupation number with the LPF control 
 {is about one third, two fifths and one half of that of cold damping with a band-pass filter, 
 that with delayed feedback,} and LQG control, respectively. 
 Thus our method has a decisive advantage to reach the absolute ground state.
 \end{abstract}
 
 \maketitle

 \section{Introduction}

 Advances in cooling techniques have played a pivotal role in the development of physics. 
 Liquefying helium~\cite{KN...11.168K} led to the discovery of superconductivity~\cite{1910KNAB...13.1274K}.
 Evaporative cooling~\cite{1996AAMOP..37..181K} %, which is a cooling technique for cold atoms, 
 brought about Bose--Einstein condensation of gaseous atoms~\cite{doi:10.1126/science.269.5221.198,PhysRevLett.75.3969,PhysRevLett.78.985}.
 Cooling techniques also play a key role in optomechanics 
 with levitated nanoparticles~\cite{millen2020optomechanics,gonzalez2021levitodynamics}, 
 which are highly isolated in a vacuum chamber and nearly free from environmental disturbances.  
 Levitated nanoparticles offer ideal platforms for studying 
 stochastic thermodynamics~\cite{gieseler2018levitated}, 
 nonequilibrium physics~\cite{PhysRevLett.110.143604} 
 and macroscopic quantum phenomena~\cite{fein2019quantum}.
 Furthermore, levitated nanoparticles have applications in 
 sensing~\cite{PhysRevA.93.053801,hempston2017force,zhu2023nanoscale,hebestreit2018sensing},  
 accelerometry, gyroscope and gravimetry~\cite{,rademacher2020quantum}. 
 Cooling is the very first step for all of these applications.

 {One of the center-of-mass degrees} of freedom 
 of a levitated nanoparticle has been cooled down to the level of
 less than one-phonon excitation~\cite{delic2020cooling,magrini2021real,Kamba:22}, 
 which is the condition usually required for ground-state cooling. 
 %Simultaneous cooling of all the translational degrees of freedom has also been endeavored, 
 %although ground-state cooling is not yet achieved. 
 %For example, s
 Simultaneous cooling of all three translational degrees of freedom of the center of mass has been
 achieved~\cite{PhysRevLett.122.123601,PhysRevApplied.22.024010}.
 Simultaneous cooling of rotational/librational degrees of freedom 
 as well as those of the center of mass has successfully been achieved~\cite{kamba2023nanoscale,pontin2023simultaneous}.
 The two widely used ground-state cooling methods are
 cavity cooling~\cite{delic2020cooling,Piotrowski:2022qda} 
 and measurement-based feedback cooling~\cite{magrini2021real,Kamba:22,tebbenjohanns2021quantum}. 
 The latter is classified in terms of feedback protocols.
 The measurement-based feedback protocols with which the ground-state cooling has been achieved 
 are cold damping \cite{Kamba:22,tebbenjohanns2021quantum} and 
 linear--quadratic--Gaussian (LQG) control \cite{magrini2021real}.

 When the detection efficiency is $\eta\, (\leqslant 1)$, 
 the phonon occupation number is fundamentally bounded from below by 
 $\frac{1}{2}(\frac{1}{\sqrt{\eta}} -1)$~\cite{doherty2012quantum,bowen2015quantum}. 
{The conventional methods can achieve this limit in the limit of weak measurement. 
 However, if the condition of weak measurement is not met,  
 the achievable phonon occupation number is above}
 this fundamental lower bound. 
 In fact, the phonon occupation number achievable via the LQG control
 reported in Ref.~\cite{PhysRevA.60.2700} 
 depends on the measurement strength.

 In this paper, we propose a feedback-cooling method 
 based on a low-pass filter in processing the measurement signal, 
 which we shall call the LPF feedback.
 We obtain the fundamental lower bound on energy achievable by the LPF feedback. 
 For the sake of comparison, 
 we also investigate the fundamental energy {bounds} 
 achievable by {cold damping with a band-pass filter and that with delayed feedback}
 for a finite measurement strength. 
 We find that the energy achievable by the LPF feedback is the lowest,
 followed by the LQG control, {cold damping with delayed feedback,
 and then by cold damping with a band-pass filter}. 
 The differences in energy between our method and the other {three} 
 increase 
 dramatically with increasing the detection efficiency and the measurement strength.

 The rest of this paper is organized as follows.
 In Section \ref{sec:quantum_trajectory}, 
 the quantum trajectory method is applied to a trapped particle under continuous measurement and feedback control 
 to obtain the equations of motion that are used in Sec.~\ref{sec:LPF} and Appendix \ref{sec:CD}.
 In Section \ref{sec:LPF}, we propose the LPF feedback
 and obtain the lowest achievable energy.
 In Section \ref{sec:Discussion}, we compare the lowest achievable energy 
 \majrev{and the highest achievable purity}
 of our LPF feedback with the LQG control{, cold damping with a band-pass filter 
 and that with delayed feedback}.
 \majrev{In Section \ref{sec:relation-btw-theory-expe}, 
 we clarify the correspondence between 
 the parameters of measurement strength and those available in experiments.}
 In Section \ref{sec:conclusion}, we conclude this paper.
 In Appendix \ref{sec:CD}, we analytically calculate the lowest energy that can be achieved 
 by cold damping {with a band-pass filter}.
 {In Appendix \ref{AppLQG}, following Ref.~\cite{PhysRevA.60.2700} 
 we obtain the lowest energy achievable by the LQG control.
 In Appendix \ref{sec:CD-DaleyedFB}, 
 we numerically obtain the lowest energy achievable by cold damping with delayed feedback, 
 which is the simplest version of cold damping widely employed in practice.}

 \section{Stochastic Dynamics under Feedback Control} \label{sec:quantum_trajectory}
 
\begin{figure}[tb]
   \centering
   \includegraphics[width=0.45\textwidth,page=1]{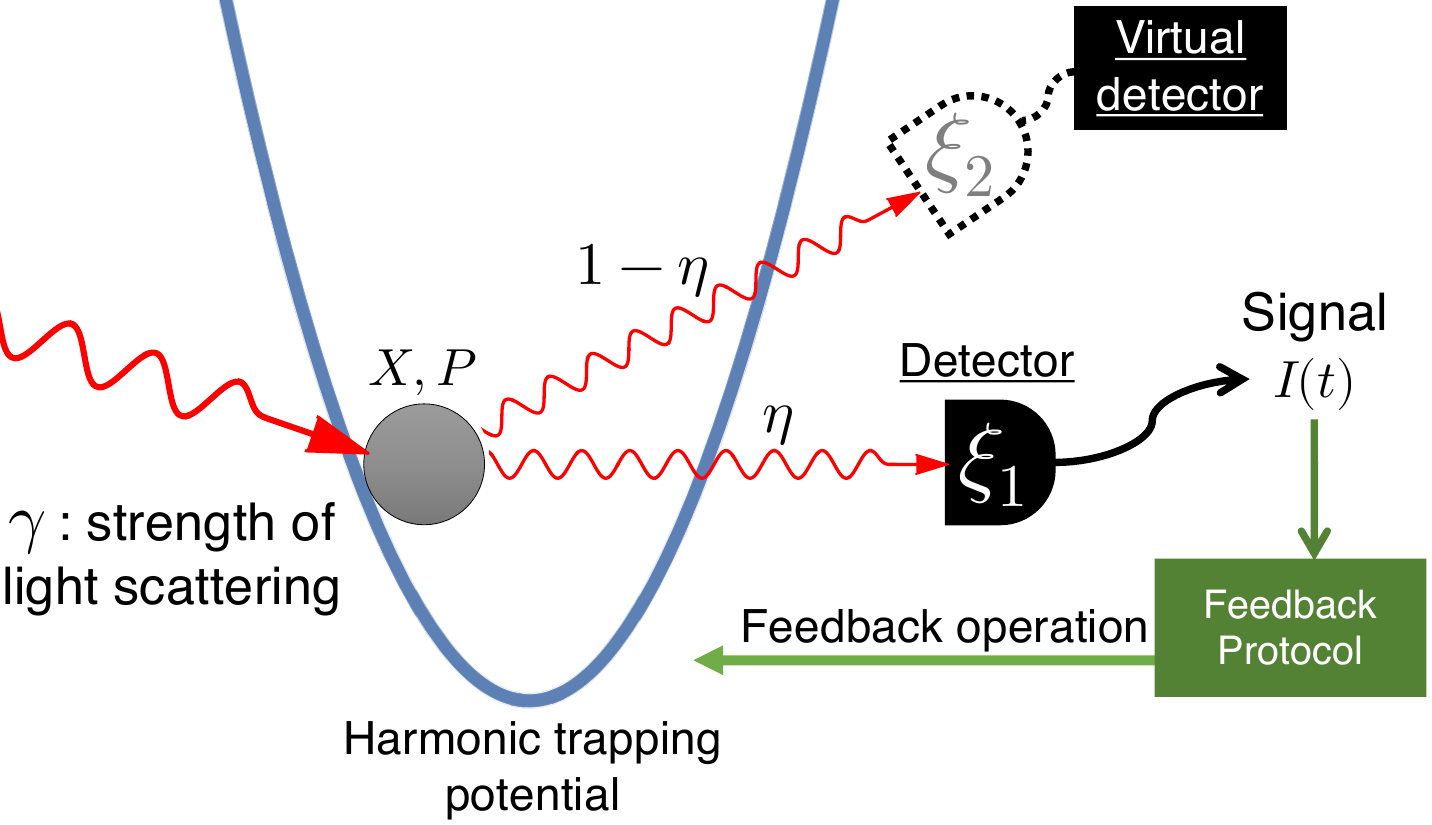}
   \caption{\label{fig:schem}
  Schematic diagram of the system under feedback control. 
  An optically trapped particle scatters an incident light with strength $\gamma$. 
  The scattered light is detected with detection efficiency $\eta$ and
  the detected light produces a measurement outcome $\xi_1$. 
  The undetection of scattered light is modeled 
  as \textit{virtual} detection with detection efficiency $1-\eta$ and a measurement outcome $\xi_2$ 
  which is not fed back to the feedback control system.
  }
 \end{figure}

 In this section, we employ the quantum trajectory method 
 to derive the equations of motion of a particle optically which is trapped 
 in a vacuum under continuous measurement and feedback control.
 As schematically illustrated in Fig.~\ref{fig:schem}, 
 an incident light is scattered by the trapped particle and the scattered light 
 is detected by a photodetector, 
 where the photocurrent provides information about the position of the particle.
 We focus on the dynamics of the trapped particle in the $x$ direction.
 A feedback protocol is applied to the particle on the basis of the observed photocurrent 
 as detailed in the following sections.

 {In experiments, there are environmental effects such as 
 collisions with the background gas and blackbody radiation \cite{gonzalez2021levitodynamics}.
 However, they are secondary to the effect of the photon recoil, 
 which is the dominant factor of heating of the center of mass of a nanoparticle.
 In fact, Ref.~\cite{Kamba:22} neglects the collisions with the background gas and 
 Refs.~\cite{magrini2021real,tebbenjohanns2021quantum} take into account 
 the environmental effects by redefining the measurement efficiency.
 Since the environmental effects can thus be incorporated in 
 the measurement efficiency, 
 we ignore them in modeling the levitated nanoparticle. 
 }

 We assume that the wavefunction of the center of mass of a particle can be described by a complex Gaussian function:
 \begin{alignat}{1}
   &\psi(x,t) = C(t) \sqrt[4]{\frac{R(t)}{\pi}} \notag \\
  &\times \exp[\frac{-R(t)+\i D(t)}{2}(x-X(t))^2 + \frac{\i P(t) x}{\hbar}].
  \label{eq:comp-Gauss-wavefunction}
 \end{alignat}
 Here $X$ and $P$ are two real parameters, $R$ is positive, $D$ is real and 
 $C$ is a complex number whose absolute value is unity.
 The variables $X$ and $P$ give the expectation values of the position and the momentum, respectively. 
 {The variables $R$ and $D$ characterize (co)variances of 
 the wavefunction \eqref{eq:comp-Gauss-wavefunction}
 in $x$ and $p$ given by 
 \begin{gather}
  V_x^\psi = \frac{1}{2R}, \quad V_p^\psi = \frac{\hbar^2 (R^2 + D^2)}{2R}, \quad V_{\rm cov}^\psi = \frac{\hbar D}{2R},
  \label{eq:covariances-complex-Gaussian-wavefnc}
 \end{gather}
 where $V_x^\psi, V_p^\psi$, and $V_{\rm cov}^\psi$ are variances in $x,p$, and their symmetrized covariance, respectively.}
 Since the global phase of the wavefunction does not affect the dynamics, 
 we will ignore the factor $C$ in the following discussions.

 We first consider the time evolution governed by the Hamiltonian
 \begin{alignat}{1}
  \hat H &= {\frac{\hat p^2}{2m} + \frac{1}{2}m\omega^2 (\hat x-X_0)^2} - F_{\rm fb} \hat x \notag \\
  &=: {\hat H_0} - F_{\rm fb} \hat x,
 \end{alignat}
 {where 
 $\hat x$ and $\hat p$ represent the position operator and the momentum operator, respectively,
 that satisfy the canonical commutation relation $[\hat x, \hat p] = \i\hbar$;}
 the three terms on the right-hand side describe the kinetic energy, 
 the shifted harmonic-potential energy, 
 and the feedback force potential. 
 Here $X_0$ represents the shift of the potential caused by the feedback control 
 and $F_{\rm fb}$ represents an external feedback force
 which shifts the minimum point of the potential.
 The shift of the potential and the external feedback force are used 
 as the key control parameters for the LPF feedback 
 and cold damping \cite{PhysRevLett.96.043003,tebbenjohanns2021quantum,Kamba:22}, respectively.
 Here and henceforth, we denote operators that act on Hilbert-space vectors with circumflexes.

 To understand how the parameters in the complex Gaussian wavefunction change in time, 
 we consider an infinitesimal change of the wavefunction:
 \begin{alignat}{1}
   &\d \ln \psi = \frac{\d \psi}{\psi}= \frac{1}{\psi} \frac{-\i H \psi\, \d t}{\hbar} \notag \\
   &= \frac{\i \hbar \d t}{2m} \left\{(-R + \i D) + \left[(-R + \i D)(x-X) + \frac{\i P}{\hbar}\right]^2\right\} \notag \\
   &\qquad + \frac{-\i m\omega^2 \d t}{2\hbar} (x-X_0)^2 + \frac{\i F_{\rm fb} \d t}{\hbar} x. \label{eq:dynamical-change}
 \end{alignat}
 The dynamical change in Eq.~\eqref{eq:dynamical-change} may be interpreted as being caused
 by the changes of the real parameters $X, P, R$ and $D$ as 
 %%%%%%%%%%%%%%%%%%%%%%%%%%%%%%%%%%%%%%%%%%%%%%%%%%%%%%%%%%%%%%%%%%%%%%%%%%%%%%%%%%%%%%%%%%%%%%%%
 \begin{alignat}{1}
  \d \ln \psi &= \frac{\d C}{C} + \frac{\d R}{4R} 
  + \frac{-\d R + \i \d D}{2}(x-X)^2 \notag \\
  & + (-R + \i D)(x-X)(-\d X) + \frac{\i \d P x}{\hbar}.
 \end{alignat}
 By identifying the above two infinitesimal changes, 
 we find that the four parameters obey the following equations:
 \begin{alignat}{1}
  \d X &= \frac{P \d t}{m}, \\
  \d P &= - m\omega^2 (X - X_0) \d t + F_{\rm fb} \d t, \\
  \d R &= -\frac{2\hbar RD}{m}\d t, \\
  \d D &= \left[\frac{\hbar}{m}(R^2 - D^2) - \frac{m\omega^2}{\hbar}\right]\d t.
 \end{alignat}
 Solving these equations, we obtain the unitary dynamics of the system.

 Let us next discuss the measurement backaction which makes the state evolution nonunitary and stochastic.
 We consider a continuous position measurement with imperfect detection efficiency. 
 We measure the position of a particle by detecting light scattered by the particle 
 with detection efficiency $\eta$ ($0\leqslant \eta \leqslant 1$) and
 fail to detect the particle with probability $1-\eta$.
 To model such imperfect detection, we introduce two measurement operators 
 for successful detection and virtual detection, as  schematically illustrated in Fig.~\ref{fig:schem}.
 The measurement outcome of the latter event can, of course, be neither known nor fed back to the system.
 Nevertheless we introduce it to describe the backaction of undetected events on the particle.
 We introduce the corresponding Kraus operators as 
 \begin{alignat}{1}
  \hat M_{\xi_1} &= \sqrt[4]{\frac{\alpha\eta}{\pi}}\exp\left[-\frac{1}{2}\alpha\eta (\hat x - \xi_1)^2\right], \label{Krausop1} \\
  \hat M'_{\xi_2} &= \sqrt[4]{\frac{\alpha(1-\eta)}{\pi}}\exp\left[-\frac{1}{2}\alpha(1-\eta) (\hat x - \xi_2)^2\right].\label{Krausop2}
 \end{alignat}
 The first one corresponds to the measurement of the detected light and 
 the second one to the virtual measurement of the undetected light.
 {Here $\alpha > 0$ quantifies the effect of the light scattering 
 over a duration of $\Delta t$.}
 We take the Kraus operators to be Gaussian 
 to make the postmeasurement wavefunction closer to a complex Gaussian wavefunction 
 even if the initial wavefunction is not so.
 As the procedure of taking the continuous measurement limit~\cite{Di_si_1988}, 
 we first consider the repeated measurements at an interval of $\Delta t$ 
 and then take the limit of $\Delta t\to 0$ with $\gamma = \alpha / \Delta t$ kept constant.
 {Here $\gamma$ represents the light scattering per unit time.}

 To study the stochastic behavior of the measurement backaction, 
 we examine the probability distribution of the measurement outcome.
 The probability density distribution of obtaining a set of measurement outcomes $(\xi_1,\xi_2)$ 
 is given by
 \begin{alignat}{1}
  &\mathrm{Prob} (\xi_1,\xi_2)
  = \langle \psi| (\hat M_{\xi_1} \hat M'_{\xi_2})^\dagger \hat M_{\xi_1} \hat M'_{\xi_2} |\psi\rangle \notag \\
   &\propto \exp[ - \alpha \eta (\xi_1 - X)^2 - \alpha (1-\eta)(\xi_2 - X)^2 + \mathcal{O}(\alpha^2)], \label{eq:051401}
 \end{alignat}
 where we omit irrelevant terms that do not depend on $\xi_1$ and $\xi_2$ 
 and terms of second and higher order in $\alpha$.
 It follows from the exponent in Eq.~\eqref{eq:051401} 
 that $\xi_1$ and $\xi_2$ obey 
 normal distributions with the same mean $X$ and 
 different variances $\frac{1}{2\alpha \eta} = \frac{1}{2\gamma\eta \Delta t}$ 
 and $\frac{1}{2\alpha(1-\eta)}$, respectively.
 Therefore, they can be expressed in terms of two independent Wiener increments $\Delta W_1$ and $\Delta W_2$ as 
 \begin{alignat}{2}
  &(\xi_1 - X) \Delta t = \frac{1}{\sqrt{2\gamma \eta}}\Delta W_1,\\
  &(\xi_2 - X) \Delta t = \frac{1}{\sqrt{2\gamma (1-\eta)}}\Delta W_2. 
 \end{alignat}
 Note that the Wiener increment $\Delta W_{i}\, (i=1,2)$ is normally distributed with mean 0 and variance $\Delta t$.
 The value of $\xi_1$ correnponds to the measurement outcome and
 the obtained signal in the limit of $\Delta t \to 0$ is given by
 \begin{equation}
  I = \lim_{\Delta t \to 0} \xi_1 = X + \frac{1}{\sqrt{2\gamma\eta}}\frac{\d W_1}{\d t}.
  \label{eq:measurement-outcome}
 \end{equation}
 Note that this expression of the measurement signal is equivalent to the corresponding expressions 
 appearing 
 in Refs.~\cite{Di_si_1988,PhysRevA.60.2700,PhysRevLett.70.548}
 except for overall constant factors.
 Upon measurement, the wavefunction undergoes measurement backaction.
 Let the measurement result be $(\xi_1, \xi_2)$.
 The wavefunction after the measurement is given by
 \begin{widetext}
   \begin{alignat}{1}
  \frac{M_{\xi_1} M'_{\xi_2} \psi(x) }{\norm{\hat M_{\xi_1} \hat M'_{\xi_2} |\psi\rangle}}
   &\propto \exp[
  \frac{-R+\i D}{2}(x-X)^2 + \frac{\i P x}{\hbar}
  -\frac{1}{2}\alpha\eta(x-\xi_1)^2 - \frac{1}{2}\alpha(1-\eta)(x-\xi_2)^2].
 \end{alignat}
 \end{widetext}
 We express this change in terms of the associated changes in the real parameters $X,P,R$ and $D$, 
 and denote their postmeasurement values with the prime as $X'$, etc.
 The changes of $R$ and $D$ are  
 \begin{equation}
  R' = R + \alpha, \quad D' = D.
 \end{equation} 
 The change of $X$ is 
 \begin{alignat}{1}
  X' &= \frac{1}{1+\frac{\alpha}{R}} \left[X+\frac{\alpha}{R}(\eta \xi_1 + (1-\eta) \xi_2) \right] \notag \\
   &\approx X + \frac{\alpha}{R}[\eta \xi_1 + (1-\eta)\xi_2 - X],\label{eq042101}
 \end{alignat}
 where the second and higher-order terms of $\Delta t$ and $\alpha$ are neglected.
 Substituting the expressions of $\xi_1$ and $\xi_2$ in Eq.~\eqref{eq042101},
 we obtain 
 \begin{equation}
  X' = X + \frac{1}{R}\left(\sqrt{\frac{\gamma\eta}{2}}\Delta W_1 + \sqrt{\frac{\gamma(1-\eta)}{2}} \Delta W_2\right).
  \label{eq:032601}
 \end{equation}
 The change of $P$ can similarly be calculated as 
 \begin{equation}
  P' = P + \hbar D (X'-X). \label{eq:032602}
 \end{equation}
 By substituting Eq.~\eqref{eq:032601} into Eq.~\eqref{eq:032602}, we obtain 
 \begin{equation}
  P' = P + \frac{\hbar D}{R}\left(\sqrt{\frac{\gamma\eta}{2}}\Delta W_1 + \sqrt{\frac{\gamma(1-\eta)}{2}} \Delta W_2\right).
 \end{equation}
 Adding the above two factors and taking the limit of $\Delta t\to 0$, 
 we obtain the equations of motion of the relevant parameters under continuous measurement as follows:
 \begin{alignat}{1}
  \d X &= \frac{P}{m}\d t + \sqrt{\frac{\gamma\eta}{2R^2}}\d W_1 + \sqrt{\frac{\gamma(1-\eta)}{2R^2}} \d W_2, \\
  \d P &= - m\omega^2 (X-X_0)\d t + F_{\rm fb} \d t \notag \\ 
  &\quad + \hbar \sqrt{\frac{\gamma\eta D^2}{2R^2}}\d W_1 + \hbar \sqrt{\frac{\gamma(1-\eta)D^2}{2R^2}}\d W_2,  \\
  \frac{\d R}{\d t} &= {-} \frac{2\hbar RD}{m} + \frac{\alpha}{\Delta t} = {-} \frac{2\hbar RD}{m} + \gamma, \\
  \frac{\d D}{\d t} &= \frac{\hbar}{m}(R^2 - D^2) - \frac{m\omega^2}{\hbar}, \\
  \d Q &= X \d t +\frac{1}{\sqrt{2\gamma\eta}}\d W_1.
 \end{alignat}
 
 We note that the differential equations obeyed by $R$ and $D$ are independent of the feedback control.
 They can be combined together in terms of a new complex parameter $z = -R+\i D$ as 
 \begin{equation}
  \frac{\d}{\d t} z = \frac{\i \hbar}{m} z^2 - \gamma - \i \frac{m\omega^2}{\hbar}.
 \end{equation}
 The solution is 
 \begin{equation}
  z(t) = \frac{1+C_0 \e^{2\i\hbar z_0 t/m}}{1-C_0 \e^{2\i \hbar z_0 t/m}} z_0,
 \end{equation}
 where 
 \begin{alignat}{1}
  C_0 &= \frac{z(0)-z_0}{z(0)+z_0}, \\
  z_0 &= \frac{m\omega}{\sqrt{2}\hbar} 
  \left[ -
  \sqrt{\sqrt{1+\tilde \gamma^2} + 1}
  + \i \sqrt{\sqrt{1+\tilde \gamma^2} - 1}
  \right],\\
  \tilde \gamma &= \frac{\hbar\gamma}{m\omega^2}.
 \end{alignat}
 {Here $\tilde \gamma$ represents a dimensionless parameter 
 of $\gamma$.}
 The solution $z(t)$ converges to $z_0$ in the limit of $t \to +\infty$, 
 and therefore
 \begin{alignat}{1}
  R &\to \frac{m\omega}{\sqrt{2}\hbar} \sqrt{\sqrt{1+\tilde \gamma^2} + 1},\label{eq:032603} \\
  D &\to \frac{m\omega}{\sqrt{2}\hbar} \sqrt{\sqrt{1+\tilde \gamma^2} - 1}.\label{eq:032604}
 \end{alignat}
 In the following sections, we use these results to obtain the lowest achievable energy under feedback cooling 
 by solving the equations regarding $X$ and $P$.

 The expectation values of the position and the momentum are $X$ and $P$, respectively.
 The variance of the position is $\frac{1}{2R}$.
 The expectation value of $\hat H_0$ is calculated as
 \begin{alignat}{1}
  \langle \psi| \hat H_0 |\psi \rangle
   &= \frac{P^2}{2m}+\frac{1}{2}m\omega^2 (X-X_0)^2 \notag \\
   &\quad + \frac{\hbar^2(R^2+D^2) + m^2\omega^2}{4mR}. \label{eq011802}
 \end{alignat}
 After $R$ and $D$ are replaced by the obtained expressions on the right-hand sides 
 of Eqs.~\eqref{eq:032603} and \eqref{eq:032604}, 
 the last term in Eq.~\eqref{eq011802} becomes
 \begin{equation}
  \frac{\hbar\omega}{2} \sqrt{\frac{\sqrt{1+\tilde \gamma^2}+1}{2}}. \label{eq011801}
 \end{equation}
 In the limit of $\tilde \gamma \to 0$, 
 Eq.~\eqref{eq011801} reduces to the ground-state energy, $\frac{1}{2}\hbar\omega$.
 Thus this term gives the zero-point energy of the system under continuous measurement. 
 The fact that this term is larger than $\frac{1}{2}\hbar\omega$ is attributed 
 to the wavefunction reduction due to backaction of the position measurement.

 {While the problem under consideration includes 
 the case of imperfect detection $(\eta < 1)$ 
 which renders a quantum state into a \textit{mixed} one, 
 we have so far discussed the case of a \textit{pure} state represented by the complex Gaussian wavefunction \eqref{eq:comp-Gauss-wavefunction}.
 The reason is the following. 
 In our formalism, 
 a mixed quantum state is unraveled with respect to the outcome of the virtual measurement, 
 to which we have no access. 
 Since the unraveled pure state 
 amounts to the mixed state upon ensemble averaging, 
 we only consider the pure state here.
 }
 
 Before closing this section, we remark on the complex Gaussian wavefunction.
 Since the Kraus operators introduced in Eqs.~\eqref{Krausop1} and \eqref{Krausop2} are Gaussian, 
 the wavefunction tends to become complex Gaussian.
 The time evolution generated by the harmonic-oscillator Hamiltonian keeps the wavefunction complex-Gaussian.
 Thus the complex Gaussian wavefunction is a natural assumption.

 \section{Cooling by Low-Pass-Filter Feedback}\label{sec:LPF}
 
\begin{figure}[tb]
   \centering
   \includegraphics[width=0.45\textwidth,page=2]{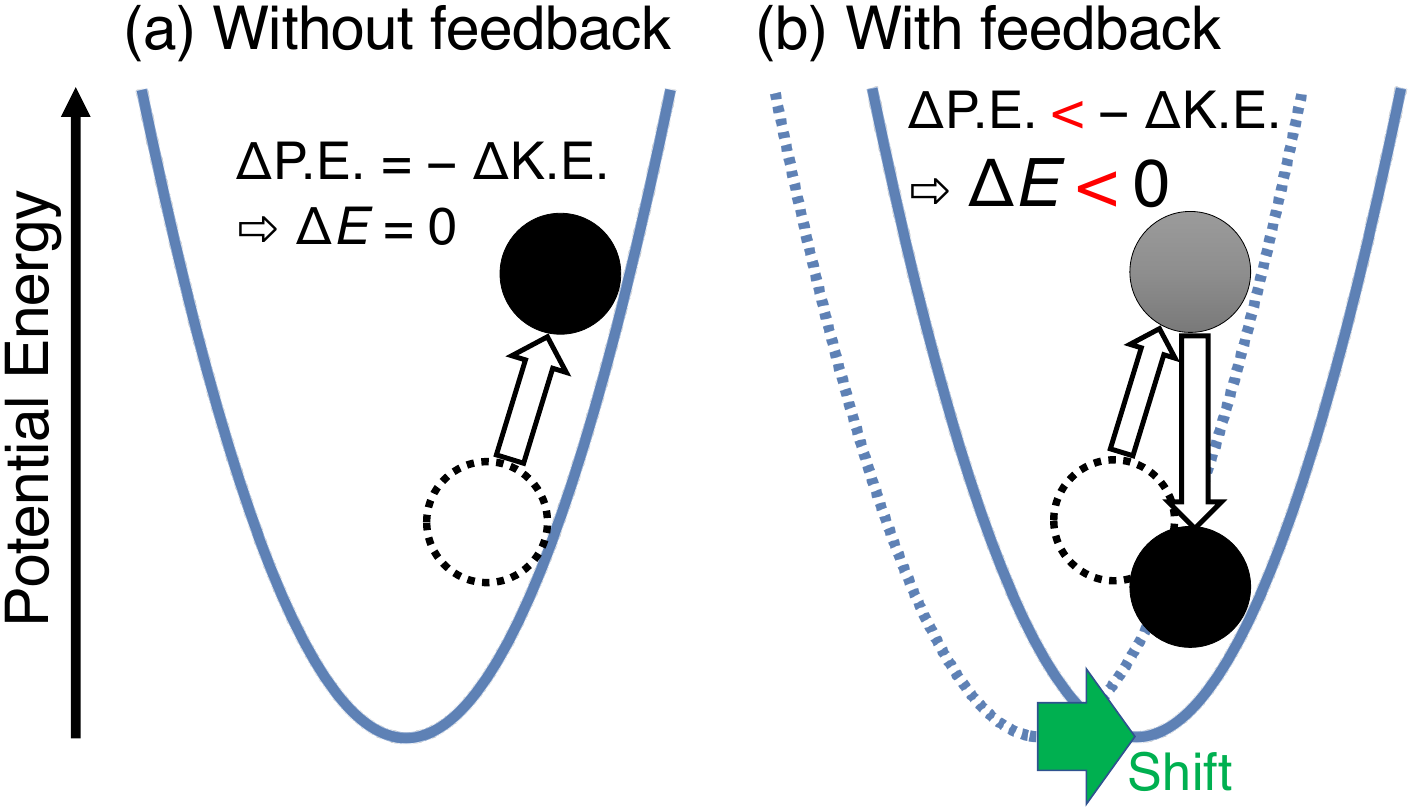}
   \caption{\label{fig:mechanism}
  Mechanism of cooling by the LPF feedback, 
  where the center of the potential is made to follow the position of the particle. 
  (a) In the absence of the LPF feedback, 
  the change of the kinetic energy (K.E.) is exactly compensated 
  for by that of the potential energy (P.E.). 
  In this case, the total energy of the system remains unchanged ($\Delta E = 0$).
  (b) Under the LPF feedback, the center of the potential is shifted on the basis of the measurement outcome
  so that the potential energy of the particle is removed. 
  Hence the total energy decreases in time and the particle is cooled ($\Delta E < 0$).
  }
 \end{figure}

 We propose a method of feedback cooling by utilizing a low-pass filter, 
 which we shall refer to the LPF feedback.
 The key idea of the LPF feedback is to shift the center of the trapping potential 
 on the basis of the position measurement of a particle 
 so that the potential minimum follows the position of the particle (see Fig.~\ref{fig:mechanism}).
 In this way, we can eliminate the potential energy of the particle and thereby cool it. 
 Moreover, we can also reduce the kinetic energy of the particle 
 by introducing a delay time between the position measurement and the action of the feedback.
 Let us discuss how a particle can be cooled under this feedback control,
 where for simplicity, we ignore the backaction from an incident light.
 We note that without feedback control 
 the gain of the potential energy is equal to the loss of the kinetic energy
 due to the conservation of the total energy (Fig.~\ref{fig:mechanism} (a)).
 By making the center of the potential follow the particle, 
 we can remove the potential energy of the particle and hence reduce its total energy. 
 Consequently, the kinetic energy of the particle decreases as the particle moves up the potential. 
 By removing this increase in the potential energy 
 by shifting the potential to the estimated position of the particle, 
 we can efficiently cool the particle as detailed below.

\begin{figure}[tb]
   \centering
   \includegraphics[width=0.45\textwidth,page=3]{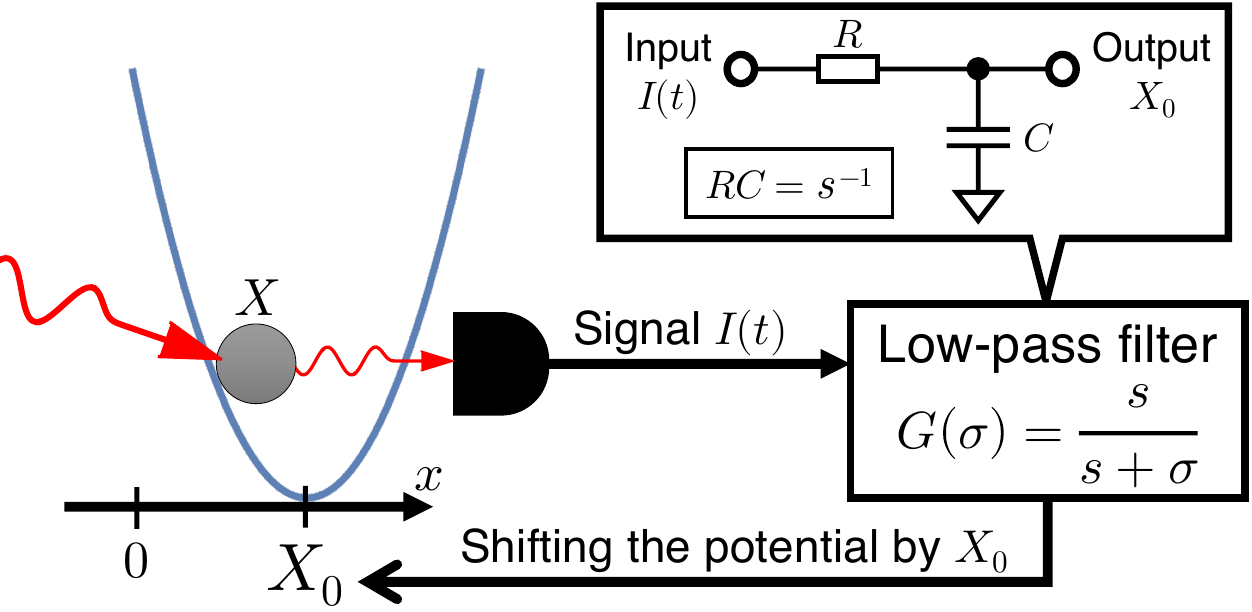}
   \caption{\label{fig:ex_imple} 
  Schematic diagram of the LPF feedback. 
  The signal, which carries information about the position of the particle, 
  is processed by a low-pass filter characterized by the transfer function $G(\sigma)$.
  Then the potential is shifted on the basis of the filtered signal. 
  The low-pass filter is comprised of a resistor and a capacitor (top right).
  We note that in a real experiment additional processes are needed  
  such as the processing of the raw signal before the low-pass filtration and
  an appropriate amplification of the output signal from the low-pass filter.
  }
 \end{figure}

 To make the center of the potential follow the position of the particle, 
 we have to estimate from the measurement signal 
 an appropriately time-averaged position of the particle 
 to which the center of the potential is shifted.
 This is because the measured position at each instant of time fluctuates randomly 
 due to quantum fluctuations and the measurement imprecision. 
 To smooth out such fluctuations while placing a larger weight on the more recent position of the particle, 
 we propose utilizing a low-pass filter.
 According to this scheme, the position of the particle is estimated to be 
 \begin{alignat}{1}
  X_{\rm pre}(t) &= s \int_{t_0}^t I(t') \e^{-s(t-t')} \d t' \notag \\
   &= s \int_{t_0}^t \left[
  X(t') \d t' +\frac{1}{\sqrt{2\gamma\eta}}\d W_1(t')
  \right] \e^{-s(t-t')}, \label{eq:032605}
 \end{alignat}
 where $X_{\rm pre}$ is the estimated position, 
 $t_0$ is the time when the feedback control is switched on 
 and $s > 0$ is the parameter of the low-pass filter which plays a role of the cut-off frequency.
 The exponential factor $\e^{-s(t-t')}$ in the integrand is introduced 
 to give a larger weight on a more recent measurement outcome.
 Additionally, the integration mitigates random noise  
 that arises from the Wiener increment in the photocurrent.
 From Eq.~\eqref{eq:032605}, a differential equation of $X_{\rm pre}$ reads
 \begin{equation}
  \d X_{\rm pre} = s(X-X_{\rm pre})\d t + \frac{s}{\sqrt{2\gamma\eta}}\d W_1. \label{eq042102}
 \end{equation}
 Substituting $X_{\rm pre}$ for the center of the potential $X_0$,
 we obtain the equations of motion under the LPF feedback control:
 \begin{alignat}{1}
  \d X &= \frac{P}{m}\d t + \sqrt{\frac{\gamma\eta}{2R^2}}\d W_1 + \sqrt{\frac{\gamma(1-\eta)}{2R^2}} \d W_2; 
  \label{eq:EoM-X} \\
  \d P &= - m\omega^2 (X-X_{\rm pre})\d t \notag \\
  &\quad + \hbar \sqrt{\frac{\gamma\eta D^2}{2R^2}}\d W_1 + \hbar \sqrt{\frac{\gamma(1-\eta)D^2}{2R^2}}\d W_2.
  \label{eq042103}
 \end{alignat} 
 {
 To facilitate an implementation of the LPF feedback,
 we note that}
 the low-pass filtration in Eq.~\eqref{eq:032605} 
 is characterized by the transfer function
 \begin{equation}
  G(\sigma) = \frac{\mathcal{L}[X_{\rm pre}](\sigma)}{\mathcal{L}[I](\sigma)} = \frac{s}{s+\sigma},
 \end{equation}
 where $\mathcal{L}[f](\sigma)$ is the Laplace transform of a function $f(t)$ defined by 
 \begin{equation}
  \mathcal{L}[f](\sigma) = \int_0^\infty f(\tau) \e^{-\sigma \tau} \d \tau.
 \end{equation}
 The low-pass filter in Eq.~\eqref{eq:032605} can easily be implemented 
 with an RC circuit as illustrated in the top-right inset in Fig.~\ref{fig:ex_imple}.

 The three equations \eqref{eq042102}--\eqref{eq042103} can be 
 combined to give the following vector equation:
 \begin{equation}
  \d \bm X = A \bm X \d t + \bm b_1 \d W_1 + \bm b_2 \d W_2, \label{eq011601}
 \end{equation}
 where 
 \begin{gather}
  \bm X := \begin{bmatrix}
  X \\P \\ X_{\rm pre}
   \end{bmatrix}, \quad 
  A := \begin{bmatrix}
     0 & m^{-1} & 0 \\ -m\omega^2 & 0 & m\omega^2 \\
  s & 0 & -s
 \end{bmatrix},\\
  \bm b_1 := \begin{bmatrix}
  \displaystyle \sqrt{\frac{\gamma\eta}{2 R^2}} \\[7pt]
  \displaystyle \hbar \sqrt{\frac{\gamma\eta D^2}{2R^2}} \\[7pt]
  \displaystyle \frac{s}{\sqrt{2 \gamma \eta}}
   \end{bmatrix}, \quad 
  \bm b_2 := \begin{bmatrix}
  \displaystyle \sqrt{\frac{\gamma(1-\eta)}{2 R^2}} \\[7pt] 
  \displaystyle \hbar \sqrt{\frac{\gamma(1-\eta) D^2}{2R^2}} \\[7pt]
       0
   \end{bmatrix}.
 \end{gather}
 To diagonalize the matrix $A$, we introduce a matrix $Q$ defined as 
 \begin{equation}
  Q = 
   \begin{bmatrix}
       1 & \displaystyle\frac{s-\sqrt{s^2 - 4\omega^2}}{2s} &\displaystyle\frac{s+\sqrt{s^2 - 4\omega^2}}{2s} \\[7pt]
       0 & \displaystyle-\frac{m\omega^2}{s} & \displaystyle-\frac{m\omega^2}{s} \\[7pt]
       1 & 1 & 1
   \end{bmatrix}.
 \end{equation}
 We assume 
 \begin{equation}
  s \neq 2\omega, \label{eq:033101}
 \end{equation}
 so that {the matrix $A$ is diagonalizable}. 
 This assumption holds true for the case of our interest 
 as discussed at the end of this section.
 By multiplying $A$ from left by $Q^{-1}$ and from right by $Q$, 
 we can diagonalize $A$:
 \begin{alignat}{1}
  A'&:= Q^{-1} A Q=
   \begin{bmatrix}
       \lambda_1 && \\ & \lambda_2 & \\ && \lambda_3
   \end{bmatrix}, 
 \end{alignat}
 where 
 \begin{alignat}{1}
   \lambda_1 &= 0, \\
   \lambda_2 &= \dfrac{-s-\sqrt{s^2-4\omega^2}}{2}, \\
   \lambda_3 &= \dfrac{-s+\sqrt{s^2-4\omega^2}}{2}.
 \end{alignat}
 Multiplying both sides of Eq.~\eqref{eq011601} by $Q^{-1}$, 
 we obtain 
 \begin{equation}
  \d \bm q 
  = A' \bm q \d t + \bm b'_1 \d W_1 + \bm b'_2 \d W_2, \label{eq011602}
 \end{equation}
 where 
 \begin{alignat}{1}
  \bm q &=\begin{bmatrix}
  q_1 , q_2 , q_3
   \end{bmatrix}^\top 
  := Q^{-1} \bm X, \label{eq011603}
  \\
  \bm b'_1 &=
     \begin{bmatrix}
  b'_{1,1} , b'_{1,2} , b'_{1,3}
     \end{bmatrix}^\top
  :=
  Q^{-1} \bm b_1, \\
  \bm b'_2 &=
     \begin{bmatrix}
  b'_{2,1} , b'_{2,2} , b'_{2,3}
     \end{bmatrix}^\top
  := Q^{-1} \bm b_2.
 \end{alignat}
 Since $A'$ is diagonal, Eq.~\eqref{eq011602} is a set of independent differential equations.
 
 We consider a stochastic differential equation of $x_t$, 
 \begin{equation}
  \d x_t = \lambda x_t \d t + \sigma_1 \d W_{1,t} + \sigma_2 \d W_{2,t}, \label{eq:040401}
 \end{equation}
 where $\lambda, \sigma_1$ and $\sigma_2$ are constants, and 
 $\d W_{1,t}$ and $\d W_{2,t}$ are independent Wiener increments.
 The solution of Eq.~\eqref{eq:040401} is given by 
 \begin{equation}
  x_t = x_{t=0} \e^{\lambda t} + \int_0^t \left[
         \sigma_1 \e^{\lambda(t-s)} \d W_{1,s} + \sigma_2 \e^{\lambda(t-s)} \d W_{2,s} 
         \label{eq012001}
  \right].
 \end{equation}
 In a similar manner, we obtain the solution of Eq.~\eqref{eq011602} as 
 \begin{alignat}{1}
  q_i(t) &= q_i(t=0)\e^{\lambda_i t} \notag \\
  &\quad+ \int_0^t \e^{\lambda_i(t-t')}\left[
  b'_{1,i} \d W_{1,t'} + b'_{2,i} \d W_{2,t'}
  \right].
 \end{alignat}

 Let us next discuss the energy of the system  which is given by
 \begin{equation}
  E = \frac{P^2}{2m}+\frac{1}{2}m\omega^2 (X-X_{\rm pre})^2 + 
  \frac{\hbar\omega}{2} \sqrt{\frac{\sqrt{1+\tilde \gamma^2}+1}{2}}. \label{eq011604}
 \end{equation}
 Here $X_0$ in Eq.~\eqref{eq011802} is replaced by $X_{\rm pre}$ and 
 the last term on the right-hand side gives the energy of the squeezed vacuum due to the measurement backaction 
 (see Eq.~\eqref{eq011801}). 
 We use Eq.~\eqref{eq011603} to express Eq.~\eqref{eq011604} in terms of $q_i$'s.
 Then the sum of the first two terms on the right-hand side of Eq.~\eqref{eq011604} is given by 
 \begin{alignat}{1}
  &\frac{m \omega ^2}{4 s^2} \bigg[q_2^2 s \left(s+\sqrt{s^2-4 \omega ^2}\right) \notag \\
  &\qquad + q_3^2 s \left(s-\sqrt{s^2-4 \omega ^2}\right)
  +8 q_2 q_3 \omega ^2\bigg].
 \end{alignat}
 The expectation value of $q_i q_j \,(i,j=2,3)$ approaches a stationary value as
 \begin{alignat}{1}
  \mathbb{E}\left[\lim_{t\to + \infty} q_i(t) q_j(t)\right]
  = - \frac{c_{1,i}c_{1,j}+ c_{2,i}c_{2,j} }{\lambda_i + \lambda_j}.
 \end{alignat}
 Here, $\mathbb{E}$ denotes the average over the Wiener processes $W_1$ and $W_2$, 
 and we employ the It\^o isometry~\cite{oksendal2013stochastic}:
 \begin{alignat}{1}
  &\mathbb{E} \left[\int_0^t f(\tau) \d W_{i,\tau} \int_0^t g(\tau')\d W_{j,\tau'}\right] \notag \\
  &\quad = \delta_{ij} \int_0^t f(\tau) g(\tau) \d \tau,\quad (i,j=1,2). 
  \label{eq012004}
 \end{alignat}
 The value of the energy in the long-time limit %after a prolonged activation of the feedback control 
 is given as follows:
 \begin{equation}
  \mathbb{E} \left[\lim_{t\to +\infty} E(t)\right] =
  \frac{\hbar \omega}{8} 
  \left[\tilde s \left(\frac{2}{\tilde \gamma \eta}+\tilde \gamma\right)
  +\frac{2 \tilde \gamma}{\tilde s}\right]. \label{eq:032608}
 \end{equation}
 Here $\tilde s = \frac{s}{\omega}$ is a dimensionless parameter of $s$.
 Minimizing the right-hand side of Eq.~\eqref{eq:032608} with respect to $\tilde s > 0$, 
 which is controllable by modulating the low-pass filter, 
 we obtain 
 \begin{equation}
  \mathbb{E} \left[\lim_{t\to +\infty} E(t)\right] \geqslant 
  \frac{\hbar\omega}{2}\sqrt{\frac{1}{\eta} + \frac{\tilde\gamma^2}{2}}
  =: \hbar \omega \left(n_{\rm LPF} + \frac{1}{2}\right), \label{mainresult_LPF}
 \end{equation}
 where the equality holds when 
 \begin{equation}
  \tilde s 
  = \frac{\tilde \gamma}{\sqrt{\dfrac{1}{\eta} + \dfrac{\tilde \gamma^2}{2}}}, 
  \label{eq:033001}
 \end{equation}
 and $n_{\rm LPF}$ is given by 
 \begin{equation}
  n_{\rm LPF} = \frac{1}{2} \left(\sqrt{\dfrac{1}{\eta} + \dfrac{\tilde \gamma^2}{2}} - 1 \right).
  \label{eq042104}
 \end{equation}
 Equation \eqref{eq042104} indicates that 
 the achievable occupation number approaches 
 the fundamental bound $\frac{1}{2} (\frac{1}{\sqrt{\eta}}-1)$ 
 in the limit of $\tilde \gamma \to 0$ and 
 increases with increasing the strength of the incident light $\tilde \gamma$.
 Equation~\eqref{mainresult_LPF} is the main result of this section.

 Let us finally discuss the assumption of Eq.~\eqref{eq:033101}, 
 which was made in introducing the matrix $Q^{-1}$.
 In fact, $s\neq 2\omega$ holds when $\tilde s$ is set to the optimal value 
 given in Eq.~\eqref{eq:033001}, 
 since for $s=2\omega$ (i.e., $\tilde s = 2$) 
 Eq.~\eqref{eq:033001} does not hold for any $\tilde \gamma$ and $\eta$
 because $0\leqslant \eta \leqslant 1$.

 \section{Comparison with Other Methods}\label{sec:Discussion}

 In this section, we compare the LPF feedback with other methods 
 which {can be} employed for the ground-state cooling of a levitated nanoparticle: 
 cold damping {with a band-pass filter}, 
 {cold damping with delayed feedback~\cite{PhysRevLett.96.043003,tebbenjohanns2021quantum,Kamba:22}} and 
 linear--quadratic--Gaussian (LQG) control \cite{stengel1994optimal}, 
 the {last} of which is known as an optimal feedback method.

 \subsection{Comparison in terms of the achievable energy}

 As shown in Appendix \ref{sec:CD}, 
 the lowest energy achievable by cold damping with a band-pass filter is given by 
 \begin{equation}
  E = \hbar\omega\left[\frac{1}{2\sqrt{\eta}} 
  + \frac{3 }{16\sqrt[6]{\eta}}\tilde \gamma^{2/3}
  + \frac{5 \sqrt[6]{\eta} }{16}\tilde{\gamma}^{4/3} + \mathcal{O}(\tilde{\gamma}^2)\right]
    \label{eq:CDapp},
  \end{equation}
 which is valid for $\tilde \gamma \lesssim 1$.
 The LQG control applied to feedback cooling of a harmonic oscillator 
 is studied in Ref.~\cite{PhysRevA.60.2700}. 
 In the limit of an infinite feedback gain with an external force, 
 the energy of the system achievable via the LQG control is given 
 in Appendix \ref{AppLQG} by 
 \begin{equation}
  E = \hbar \omega
  \left[\frac{1}{2\sqrt{\eta}} \sqrt{\frac{\sqrt{1+\eta \tilde \gamma^2} + 1}{2}}
  + \frac{\tilde \gamma}{2(\sqrt{1+\eta \tilde \gamma^2}+1)}
  \right]. \label{eq:LQG}
 \end{equation}
 \majrev{For the lowest energy achievable 
 by feedback techniques other than LPF feedback, 
 we do not use the energy with respect to the potential center 
 which is shifted by the feedback operation
 but we employ the energy with respect to the fixed origin given by 
 \begin{equation}
  E = \frac{P^2}{2m}+\frac{1}{2}m\omega^2 X^2 + 
  \frac{\hbar\omega}{2} \sqrt{\frac{\sqrt{1+\tilde \gamma^2}+1}{2}}.
  \label{eq:energy-wrt-fixed-origin}
 \end{equation}
 This is because the conventional feedback techniques
 are designed to minimize the energy 
 given by Eq.~\eqref{eq:energy-wrt-fixed-origin};
 the adoption of the shifted energy would 
 make the achievable energy higher. 
 }

\begin{figure}[tb]
   \centering
   \includegraphics[width=0.45\textwidth]{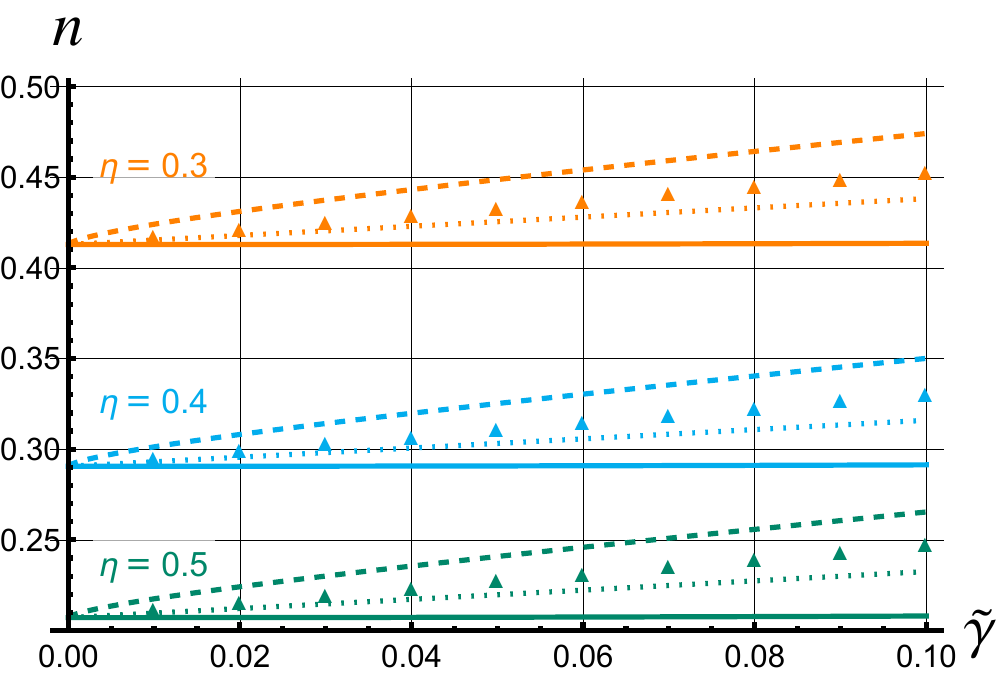}
   \caption{\label{fig:compar} {
  Phonon occupation number $n=\frac{E}{\hbar\omega} - \frac{1}{2}$ of achievable energies 
  under LPF feedback (Eq.~\eqref{mainresult_LPF}, solid curves), 
  cold damping with a band-pass filter (Eq.~\eqref{eq:CDapp}, dashed curves), 
  LQG control (Eq.~\eqref{eq:LQG}, dotted curves) 
  and cold damping with delayed feedback (numerical simulation (see Appendix \ref{sec:CD-DaleyedFB}), triangles)
  for $\eta=0.3\text{ (orange)},0.4\text{ (blue)}$ and $0.5\text{ (dark green)}$.
  Every method can cool the system down to the fundamental limit 
  $\frac{1}{2}(\frac{1}{\sqrt{\eta}} -1)$~\cite{doherty2012quantum,bowen2015quantum} 
  in the limit of $\tilde \gamma \to 0$. 
  However, for small but nonzero $\tilde \gamma$, 
  the achievable energy of the LPF feedback is the lowest of the four, 
  followed by LQG control, cold damping by delayed feedback, 
  and then by cold damping with a band-pass filter.}}
 \end{figure}

 The lowest energy that is achievable for each of the {four} feedback methods is
 plotted in Fig.~\ref{fig:compar}. 
 As shown in Fig.~\ref{fig:compar}, 
 the achievable energy via the LPF feedback is the lowest, 
 followed by LQG control, {cold damping with delayed feedback,}
 and then by cold damping with a band-pass filter. 
 However, the achievable energies are only slightly different 
 for $\tilde \gamma =0.1$.

\begin{figure}[tb]
   \centering
   \includegraphics[width=0.45\textwidth]{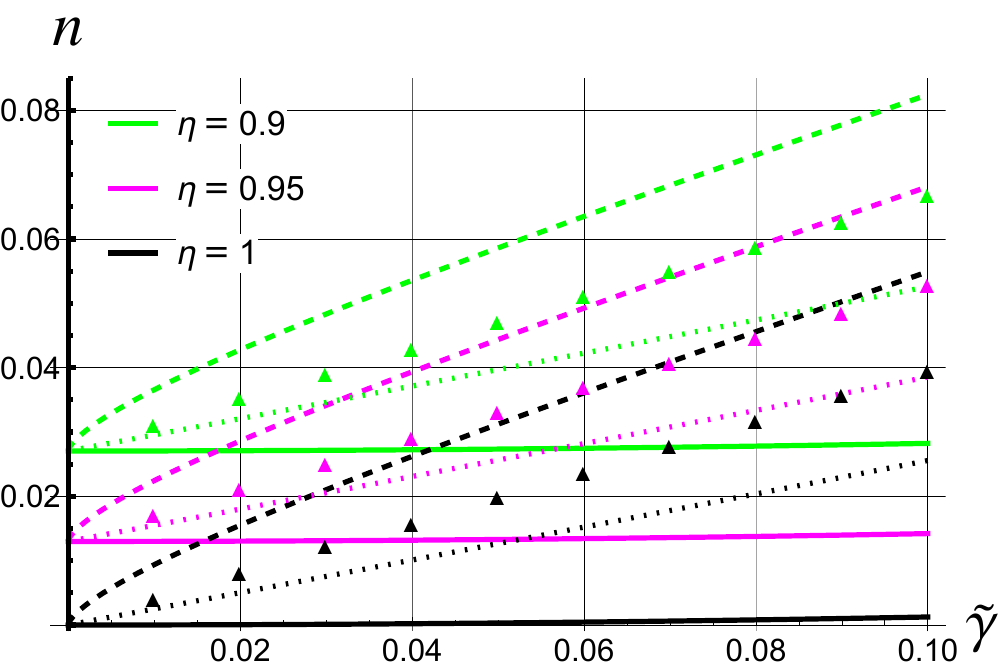}
   \caption{\label{fig:high_eff} {
  Phonon occupation number $n=\frac{E}{\hbar\omega} - \frac{1}{2}$ of achievable energies 
  under LPF feedback (Eq.~\eqref{mainresult_LPF}, solid curves), 
  cold damping with a band-pass filter (Eq.~\eqref{eq:CDapp}, dashed curves), 
  LQG control (Eq.~\eqref{eq:LQG}, dotted curves) 
  and cold damping with delayed feedback (numerical simulation (see Appendix \ref{sec:CD-DaleyedFB}), triangles)
  for $\eta=0.9\text{ (light green)},0.95\text{ (magenta)}$ and $1.0\text{ (black)}$.
  The dashed curves are calculated from Eq.~\eqref{eq:CDapp}.
  For high measurement efficiency ($\eta \approx 1$), 
  the achievable phonon occupation numbers of two versions of cold damping and LQG control increase dramatically 
  with increasing the dimensionless measurement strength $\tilde \gamma$ 
  in sharp contrast with the LPF feedback 
  where the achievable phonon occupation number stays almost constant 
  with respect to $\tilde{\gamma}$.}}
 \end{figure}

 For detection efficiencies $\eta =0.3,0.4$ and $0.5$, 
 which are typical values for the ground-state cooling experiments~\cite{magrini2021real,Kamba:22,tebbenjohanns2021quantum}, 
 the differences in the achievable energies among the three methods are small.
 However, for high detection efficiency $\eta\approx 1$, they are significant. 
 Figure~\ref{fig:high_eff} shows that for $\tilde \gamma=0.1$ with $\eta =1$ 
 the achievable phonon occupation number is about 0.025 for LQG control, 
 {0.04 for cold damping with delayed feedback,
  0.05 for cold damping with a band-pass filter}
 %dozens of times higher than that of LPF feedback 
 and almost zero for LPF feedback.
 Thus only with the LPF feedback can one reach the nearly absolute ground state.

\majrev{
 \subsection{Comparison of different feedback methods 
 in terms of the achievable purity}

 We compare the different feedback methods in terms of 
 the highest purity achievable by each method.
 We note that 
 the information available to the controller 
 about the position and the momentum could be 
 utilized for computation of purity. 
 The purity $\mathcal{P}$ can be calculated from~\cite{PhysRevA.60.2700,PhysRevA.68.012314}, 
 \begin{equation}
  \mathcal{P} = 
  \frac{\hbar}{2\sqrt{V_x^{\rm tot} V_p^{\rm tot} - (V_{\rm cov}^{\rm tot})^2}},
 \end{equation}
 where $V_x^{\rm tot}, V_p^{\rm tot}$ and $V_{\rm cov}^{\rm tot}$ are the variances in $x,p$ and 
 their covariance, respectively.
 These (co)variances will be adjusted to incorporate the controller's information.
}

 \textit{Because our LPF feedback prepares a state that is expected to be localized 
 near the estimated position $X_{\rm pre}(t)$ which is accessible to the feedback controller,}
 we evaluate the purity by using the (co)variances of the state given by
 \begin{equation}
  \hat \rho := \lim_{t\to \infty}\mathbb{E}[|T_{-X_{\rm pre}(t)}\psi (t) \rangle
  \langle T_{-X_{\rm pre}(t)} \psi(t)|],
  \label{eq:ensemble-state-LPF}
 \end{equation} 
 where 
 \begin{equation}
  \hat T_{-X_{\rm pre}(t)} := \exp[-\frac{\i \hat p (-X_{\rm pre (t)})}{\hbar}]
 \end{equation}
 is the operator that displaces the state by $-X_{\rm pre} (t)$.
 We note that the (co)variances for $\hat \rho$ is obtained 
 as the sum of those of the complex Gaussian wavefunction \eqref{eq:covariances-complex-Gaussian-wavefnc} 
 and those of $X-X_{\rm pre}$ and $P$:
 \begin{alignat}{1}
  V_x^{\rm tot} &= V_x^\psi + \lim_{t\to \infty} \mathbb{E}\left[(X(t)-X_{\rm pre}(t))^2\right],\\
  V_p^{\rm tot} &= V_p^\psi + \lim_{t\to \infty} \mathbb{E}\left[P(t)^2\right],\\
  V_{\rm cov}^{\rm tot} &= V_{\rm cov}^\psi + \lim_{t\to \infty} \mathbb{E}\left[(X(t)-X_{\rm pre}(t)) P(t)\right].
 \end{alignat} 
 We note that the (co)variances of $X-X_{\rm pre}$ and $P$ 
 can be obtained in the same way as in the computation of the lowest achievable energy 
 in Sec.~\ref{sec:LPF}.

\begin{figure}[tb]
   \centering
   \includegraphics[width=0.48\textwidth]{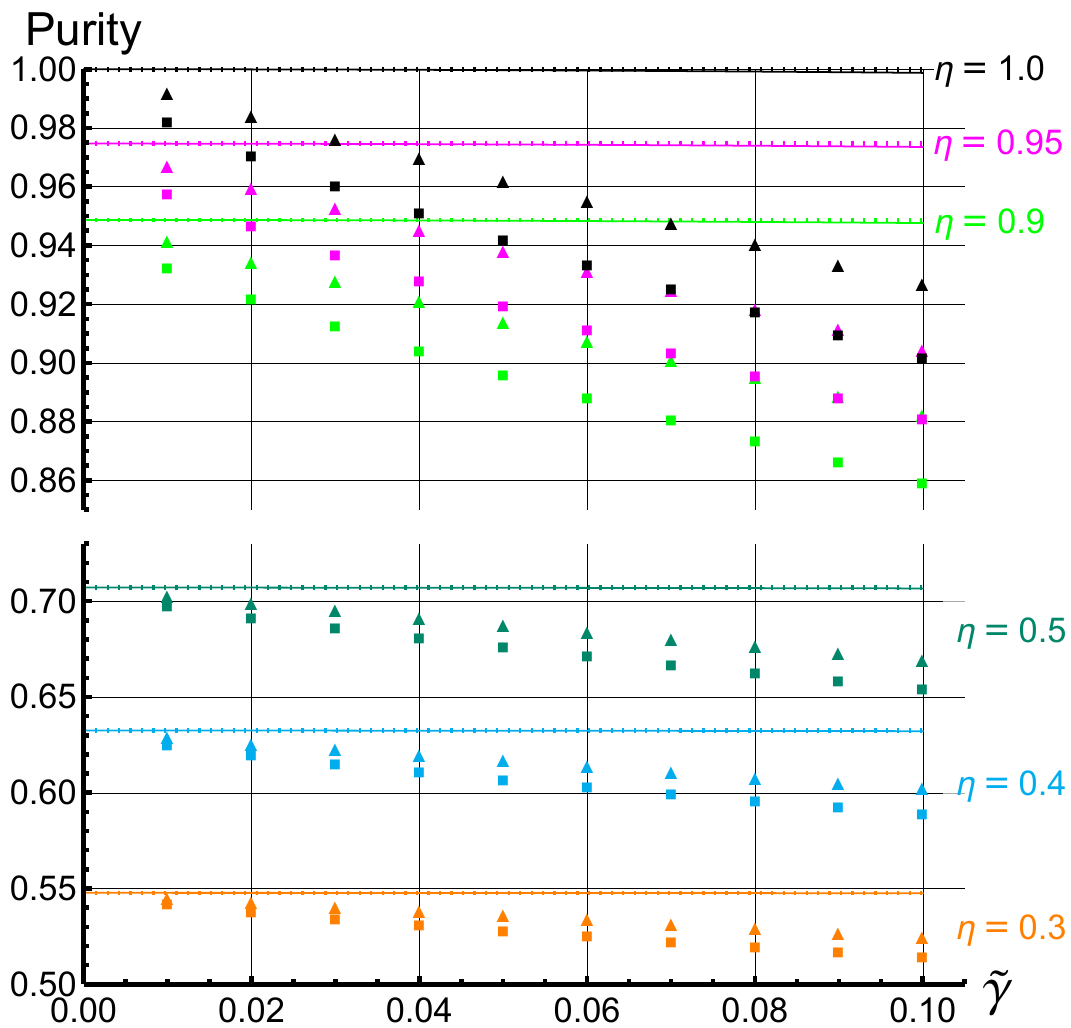}
   \caption{{ \label{fig:purity}
   Purity of the state prepared by the LPF feedback (solid curve), 
   cold damping with a band-pass filter (squares),
   cold damping with delayed feedback (triangles) and 
   the LQG control (dotted curves).  
   The upper panel shows the case of the detection efficiency 
  $\eta = 0.9$ (light green), $0.95$ (magenta), and $1$ (black),
  while lower panel shows that of 
  $\eta = 0.3$ (orange), $0.4$ (blue), and $0.5$ (dark green).
   The squares and the triangles are obtained from 
   numerical simulations of the (co)variances 
   for $\eta=0.3,0.4,0.5,0.9,0.95$ and $1$ and $\tilde \gamma = 0.01, 0.02,\dots,0.10.$
   \majrev{We find that 
   the LQG control achieves the highest purity, followed by the LPF feedback, 
   cold damping with delayed feedback, 
   and cold damping with a band-pass filter.
   The difference 
   between the LPF feedack and the LQG control is so small 
   that their curves almost overlap.}}}
 \end{figure}

\majrev{
In the LQG control, the Kalman filter provides the estimated 
position and momentum of the nanoparticle 
on the basis of the measurement outcome given in 
Eq.~\eqref{eq:measurement-outcome}. 
Incorporating this information,
we obtain the purity of the state 
for the LQG control as~\cite{PhysRevA.60.2700} 
\begin{equation}
  \mathcal{P}_{\rm LQG} = \sqrt{\eta}.
\end{equation}

For the two methods of cold damping, 
adjustment by using the controller's information is not conducted because
it does not increase the purity, 
as detailed in Appendices \ref{sec:CD} and \ref{sec:CD-DaleyedFB}.

The achievable purity computed for each method 
is plotted in Fig.~\ref{fig:purity}, 
which shows that the LQG can achieve the highest purity, 
followed by the LPF feedback, cold damping with delayed feedback, 
and then by cold damping with band-pass filter. 
Because the purity, computed by using the controller's information, 
indicates the estimation accuracy of signal filtering for each method,
the LQG control serves as the best filter for estimation.

}

\majrev{\subsection{Similarity of LPF feedback to cold damping}}

 {A similarity between the LPF control and cold damping can be found 
 in the limit of weak measurement ($\tilde \gamma \to 0$).
 From Eq.~\eqref{eq:033001}, the optimal parameter of the low-pass filter 
 approaches zero in that limit, 
 where the low-pass filtration in Eq.~\eqref{eq:032605} 
 amounts to the integration of the input signal of the measurement outcome.
 Because the oscillating and stochastic signal input into the low-pass filter 
 is suppressed, 
 the output signal will give the average of that input 
 with very small fluctuations.
 We can therefore assume a single-mode oscillation 
 \begin{equation}
  X(t) \approx \overline{X(t)} + \alpha_0 \sin (\omega t + \phi_0),
  \label{eq:FreeOscillation}
 \end{equation}
 where $\overline{X (t)}$ represents the average of $X(t)$ over a long period, 
 $\alpha_0$ and $\phi_0$ are real constants.
 For the freely oscillating nanoparticle \eqref{eq:FreeOscillation}, 
 the potential minimum is shifted by
 \begin{alignat}{1}
  X_{\rm pre} (t) &\approx \overline{X(t)} - 
  \frac{\alpha_0}{\omega} \cos (\omega t + \phi_0) \notag \\
  &\approx \overline{X(t)} - \frac{\alpha_0}{\omega^2} \frac{\d X(t)}{\d t}
  \label{eq:similarity1}
 \end{alignat} 
 The second term in Eq.~\eqref{eq:similarity1} generates 
 an effect of resistance force proportional to the negative of the velocity 
 of the nanoparticle. 
 Thus the LPF control becomes quite similar to cold damping 
 in the weak measurement limit.
 }

 \majrev{\subsection{Optimality of LQG control}}
 
 Let us finally discuss the optimality of the LQG control. 
 Although it is classified as optimal control, its
 achievable energy is higher than that of the LPF feedback.
 The reason is as follows.
 The LQG control employs a feedback protocol 
 that minimizes
 \begin{equation}
  \left\langle \frac{\hat p^2}{2m} + \frac{1}{2}m\omega^2 \hat x^2\right\rangle  , \label{eq:031301}
 \end{equation}
 which is the sum of the kinetic energy and the potential energy of the particle.
 This potential energy does not include the effect of the shift of the minimal point of the potential 
 due to the application of the feedback potential.
 On the other hand, the LPF feedback minimizes
 \begin{equation}
  \left\langle \frac{\hat p^2}{2m} + \frac{1}{2}m\omega^2 (\hat x-X_0)^2\right\rangle,
 \end{equation}
 which includes the shift of the potential \majrev{caused by 
 feedback operation}.
 This is why the achievable energy of the LQG control 
 is higher than that of the LPF feedback.

\majrev{

\section{Relation between theoretical and experimental parameters}
 \label{sec:relation-btw-theory-expe}

To facilitate comparison with experimental systems, 
we give two practical ways to compute $\tilde \gamma$ 
from experimentally available parameters.
Recall that $\tilde{\gamma}$ is the dimensionless quantity of 
the strength of light scattering $\gamma$, and 
that $\gamma$ appears in the measurement outcome \eqref{eq:measurement-outcome}. 
The spectral density (SD) of the measurement outcome is defined as 
\begin{alignat}{1}
  \mathrm{SD}[I](\omega') &= \left| \mathscr{F} [I] (\omega') \right|^2, \\
  \mathscr{F}[I](\omega') &= \frac{1}{\sqrt{2\pi}}\int I(t) \e^{-\i\omega' t} \d t.
\end{alignat}
Since the spectral density of the white noise $\frac{\d W}{\d t}$ is unity, 
\begin{equation}
  \mathrm{SD}\left[\frac{\d W}{\d t}\right] = 1,
\end{equation}
the spectral density of the measurement outcome is computed as 
\begin{equation}
  \mathrm{SD}[I]
  = \mathrm{SD}[X]
  %+ 2 \text{Re}\left[\mathscr{F} [X]^* \mathscr{F} \left[\frac{\d W}{\d t}\right]\right]
  + \frac{1}{2\gamma \eta}. \label{eq:SD1}
\end{equation}
Here the cross term between Fourier transforms of
the position of the nanoparticle and the white noise vanishes 
because they are uncorrelated. 
The first term on the right-hand side of Eq.~\eqref{eq:SD1}, 
which includes $\mathscr{F}[X]$, 
becomes negligible far from the particle's oscillation frequency~$\omega$. 
Hence, the noise floor of the measured spectrum, 
which represents the measurement precision, 
can be identified as $1/(2\gamma\eta)$. 
The measurement precision is typically expressed in units of 
$\mathsf{m}^2/\mathsf{Hz}$ or its square root $\mathsf{m}/\sqrt{\mathsf{Hz}}$.

An alternative method for determining $\tilde{\gamma}$ 
is to use the photon-recoil heating rate. 
Without feedback cooling, 
light scattering increases the center-of-mass energy of the nanoparticle as
\begin{alignat}{1}
  \mathbb{E}\left[\frac{\d}{\d t} \left(
    \frac{P^2}{2m} + \frac{1}{2} m\omega^2 X^2 \right)
  \right]
  = \frac{\hbar \gamma}{4m},\label{eq:heating-rate}
\end{alignat}
where Eqs.~\eqref{eq:EoM-X} and \eqref{eq042103} have been used.
By identifying the experimentally observed heating rate 
with Eq.~\eqref{eq:heating-rate}, 
one can extract $\gamma$ and thereby determine $\tilde{\gamma}$.

For reference, Table~\ref{tab:tildegamma} lists values of $\tilde{\gamma}$ 
for several experiments that achieved ground-state cooling. 
\begin{table*}[tb]
\caption{\label{tab:tildegamma} 
Representative values of $\tilde{\gamma}$ and the parameters 
used for their estimation in selected experiments.}
\begin{ruledtabular}
\begin{tabular}{ccccc|c}
 Experiment & Mass & Angular frequency & 
 Measurement precision & Measurement efficiency & $\tilde \gamma$ \\ \hline
 L.~Magrini \textit{et al.}~\cite{magrini2021real} &
 $2.8\times 10^{-18}\,\mathsf{kg}$ & 
 $2\pi \times 104\,\mathsf{kHz}$ & $2.0\times 10^{-14}\, \mathsf{m/\sqrt{Hz}}$ 
 & 0.34 & 0.32 \\
 F.~Tebbenjohanns \textit{et al.}~\cite{tebbenjohanns2021quantum}&
 $1\times 10^{-18}\,\mathsf{kg}$ & 
 $2\pi \times 77.6\,\mathsf{kHz}$ & $2.27\times 10^{-14}\, 
 \mathsf{m/\sqrt{Hz}}$
 & 0.24 & 1.8
\end{tabular}
\end{ruledtabular}
\end{table*}
As summarized in Table~\ref{tab:tildegamma}, 
$\tilde \gamma$ ranges from about 0.1 to 1 in the experimental systems.
}

 \section{Conclusion}\label{sec:conclusion}
 
 In this paper, 
 we employ the quantum trajectory method of feedback cooling of a trapped particle
 and propose a method of feedback control based on low-pass filtration.
 This method can achieve the lower phonon occupation number
 than LQG control, cold damping with a band-pass filter 
 and cold damping with delayed feedback, 
 \majrev{and can achieve almost the same purity as the LQG control.} 
 For parameters $\eta =0.3$ and $\tilde \gamma = 0.1$, 
 the differences in achievable energies between them are slight. 
 However, in future experiments with a significantly higher efficiency, 
 these differences become pronounced. 
 Under such situations, 
 only the LPF feedback proposed in this paper can reach the nearly absolute ground state.
 %will be essential.

 \begin{acknowledgements}
 The authors would like to thank Kiyotaka Aikawa and Mitsuyoshi Kamba 
 for useful comments from an experimental viewpoint, 
 and Koki Shiraishi and Masaya Nakagawa for fruitful theoretical discussions.
 This work was supported by KAKENHI Grant No.~JP22H01152 from the Japan Society for the Promotion of Science. 
 We gratefully acknowledge the support from the CREST program ``Quantum Frontiers" 
 (Grant No.~JPMJCR23I1) by the Japan Science and Technology Agency.
 \end{acknowledgements}

 \appendix

 \section{Achievable Minimum Energy by Cold Damping with a Band-Pass Filter}\label{sec:CD}

 For the purpose of comparison of our result discussed in Sec.~\ref{sec:Discussion}, 
 we examine cold damping with a band-pass filter.
 %\cite{PhysRevLett.96.043003,tebbenjohanns2021quantum,Kamba:22}, 
 %which is among the standard methods of feedback cooling of a trapped particle. 
 The feedback force applied to a trapped particle is 
  proportional to the velocity of the particle 
 and the direction of the force is opposite to it. 
 Thus the particle is decelerated and cooled.
 %The cold damping is experimentally implemented as follows~\cite{PhysRevLett.96.043003,tebbenjohanns2021quantum,Kamba:22}:
 {Cold damping with a band-pass filter, 
 which we note is neither the only nor the main method of cold damping, 
 is implemented as follows:}
 a band-pass filter allows the photocurrent 
 to pass only at a given angular frequency $\omega$; 
 the filtered signal is shifted by $\frac{\pi}{2}$ 
 to obtain the signal proportional to the velocity of the particle, 
 and a feedback force is applied on the basis of the processed signal. 
 Here it is assumed that the particle oscillates at the frequency $\omega$ of the trapping potential.

 We extract the signal with frequency $\omega$ from the photocurrent as 
 \begin{alignat}{1}
  I^{\rm (est)}(t) &:= {\rm Re} [Z(t) \e^{\i\omega t}] 
  = \frac{1}{2}\left(Z(t) \e^{\i\omega t} + Z(t)^* \e^{-\i\omega t}\right), \\
  Z(t) &:= 2s \int^t I(t') \exp[-\i\omega t' -s(t-t')] \d t' . \label{eq011901}
 \end{alignat}
 Here $I^{\rm (est)}(t)$ is the filtered signal,
 $Z(t)$ is a complex amplitude of the signal at frequency $\omega$
 and $s$ is a cut-off frequency of the band-pass filter.
 Note that $I^{\rm (est)}(t)$ estimates the position of the particle.
 Under the assumption that $Z(t)$ varies much slower than $\e^{i\omega t}$,
 the estimated velocity of the particle is given by
 \begin{alignat}{1}
  \frac{\d}{\d t}{\rm Re} \left[Z\e^{\i\omega t}\right] 
   &\approx {\rm Re} \left[\i\omega Z\e^{\i\omega t}\right] \notag \\
  &= \frac{\i\omega}{2}\left(Z \e^{\i\omega t} - Z^* \e^{-\i\omega t}\right) 
  =: v_{\rm est}.
 \end{alignat}
 Then the following feedback force is applied to the particle:
 \begin{equation}
  F_{\rm fb} =  
  -g\frac{\i\omega}{2}\left(Z \e^{\i\omega t} - Z^* \e^{-\i\omega t}\right). \label{eq:032606}
 \end{equation}
 Here $g$ represents the gain of the feedback control.
 Differentiating Eq.~\eqref{eq011901}, we obtain 
 \begin{equation}
  \d Z = 2s \e^{-\i\omega t} \left(X \d t + \frac{\d W_1}{\sqrt{2\gamma\eta}}\right) 
  + (-s) Z \d  t.\label{eq:032607}
 \end{equation}

 The time dependence of the coefficients on the right-hand sides of 
 Eqs.~\eqref{eq:032606} and \eqref{eq:032607} can be eliminated 
 by introducing a new variable $Y=\e^{i\omega t} Z$.
 The equations of motion are then given as follows:
 \begin{alignat}{1}
  \d X &= \frac{P}{m}\d t + \sqrt{\frac{\gamma\eta}{2R^2}}\d W_1 
  + \sqrt{\frac{\gamma(1-\eta)}{2R^2}} \d W_2, \\
  \d P &= - m\omega^2 X\d t -g\frac{\i\omega}{2}\left(Y - Y^*\right) \d t \notag \\
  &\quad + \hbar \sqrt{\frac{\gamma\eta D^2}{2R^2}}\d W_1 
  + \hbar \sqrt{\frac{\gamma(1-\eta)D^2}{2R^2}}\d W_2,  \\
  \d Y &= 2s X \d t + (+\i\omega -s) Y \d t 
  + \frac{2s}{\sqrt{2\gamma\eta}}\d W_1, \\
  \d Y^* &= 2s X \d t + (-\i\omega -s) Y^* \d t 
  + \frac{2s}{\sqrt{2\gamma\eta}}\d W_1.
 \end{alignat}
 Here $Y^*$ is the complex conjugate of $Y$.
 The above four equations can be expressed in vector form:
 \begin{alignat}{1}
  \d \bm Y &= B \bm Y \d \phi + \bm c_1 \d w_1 + \bm c_2 \d w_2, \label{eq011902}
 \end{alignat}
 where
  \begin{alignat}{1}
  \bm Y &= [Y_1,Y_2,Y_3,Y_4]^\top = \left[mX, \frac{P}{\omega}, mY, mY^*\right]^\top,  \\
  B &= \begin{bmatrix}
       0 & 1 & 0 & 0 \\
  -1 & 0 & -\i \tilde g & \i \tilde g \\
       2\tilde s& 0 & \i -\tilde s & 0 \\
       2\tilde s & 0 & 0 & -\i -\tilde s
   \end{bmatrix}, \\
  \tilde g &:= \frac{g}{2m\omega}, \quad 
   \phi := \omega t,\\[3mm]
  \bm c_1 &= 
   \begin{bmatrix}
  \sqrt{\frac{m\hbar}{2\omega} \frac{\eta \tilde\gamma}{\tilde R^2}} \\
  \sqrt{\frac{m\hbar}{2\omega} \frac{\eta \tilde\gamma \tilde D^2}{\tilde R^2}} \\
  \sqrt{\frac{2m\hbar}{\omega} \frac{\tilde s^2}{\eta \tilde\gamma}} \\
  \sqrt{\frac{2m\hbar}{\omega} \frac{\tilde s^2}{\eta \tilde\gamma}} 
   \end{bmatrix},  \quad 
  \bm c_2 = 
   \begin{bmatrix}
  \sqrt{\frac{m\hbar}{2\omega} \frac{(1-\eta) \tilde\gamma}{\tilde R^2}} \\
  \sqrt{\frac{m\hbar}{2\omega} \frac{(1-\eta) \tilde\gamma \tilde D^2}{\tilde R^2}} \\
       0\\ 0
   \end{bmatrix},\\
  \tilde R &= \frac{\hbar R}{m\omega},\quad  \tilde D = \frac{\hbar D}{m\omega},\\
  \d w_1 &= \sqrt{\omega} \d W_1, \quad \d w_2 = \sqrt{\omega} \d W_2.
 \end{alignat}
 Here dimensionless quantities $\tilde g,\phi, \tilde R, \tilde D, \d w_1$ and $\d w_2$ 
 correspond to $g, t, R, D, \d W_1$ and $\d W_2$, respectively. 
 We assume $\tilde s \neq 4 \tilde g$
 so that the matrix $B$ is diagonalizable.
 Let $V$ be a matrix whose column vectors form a basis consisting of eigenvectors of $B$.
 %Such $B$ is not unique but the following discussion is independent of its choice. 
 By multiplying $B$ from left by $V^{-1}$ and from right by $V$, 
 we can diagonalize $B$ as 
 \begin{alignat}{1}
  B'&:= V^{-1} B V =
   \begin{bmatrix}
       \begin{matrix}
         \nu_1 & \\ & \nu_2
       \end{matrix} & \mbox{\LARGE 0} \\
  \raisebox{-1.2ex}{\mbox{\LARGE 0}} & 
       \begin{matrix}
         \nu_3 & \\ & \nu_4
       \end{matrix}
   \end{bmatrix}, \\
\end{alignat}
where
\begin{alignat}{1}
   \nu_1 &= \frac{- \tilde s - \sqrt{\tilde s^2 -4- 4\sqrt{4\tilde g \tilde s - \tilde s^2}}}{2} , \label{nu1} \\
   \nu_2 &= \frac{- \tilde s + \sqrt{\tilde s^2 -4- 4\sqrt{4\tilde g \tilde s - \tilde s^2}}}{2} , \label{nu2} \\
   \nu_3 &= \frac{- \tilde s - \sqrt{\tilde s^2 -4 + 4\sqrt{4\tilde g \tilde s - \tilde s^2}}}{2} , \label{nu3} \\
   \nu_4 &= \frac{- \tilde s + \sqrt{\tilde s^2 -4 + 4\sqrt{4\tilde g \tilde s - \tilde s^2}}}{2}. \label{nu4}
 \end{alignat}
 Multiplying both sides of Eq.~\eqref{eq011902} by $V^{-1}$, 
 we obtain 
 \begin{equation}
  \d \bm p
  = B' \bm p \d \phi + \bm c'_1 \d w_1 + \bm c'_2 \d w_2, \label{eq011903}
 \end{equation}
 where 
 \begin{alignat}{1}
  \bm p &=\begin{bmatrix}
  p_1 , p_2 , p_3 , p_4
   \end{bmatrix}^\top := V^{-1} \bm Y, \label{eq011904}
  \\
  \bm c'_1 &=
     \begin{bmatrix}
  c'_{1,1} , c'_{1,2} , c'_{1,3} , c'_{1,4}
     \end{bmatrix}^\top
  :=
  V^{-1} \bm c_1, \\
  \bm c'_2 &=
     \begin{bmatrix}
  c'_{2,1} , c'_{2,2} , c'_{2,3} , c'_{2,4}
     \end{bmatrix}^\top
  := V^{-1} \bm c_2.
 \end{alignat}
 Since $B'$ is diagonal, 
 Eq.~\eqref{eq011903} represents 
 a set of four independent stochastic differential equations.
 Hence, using the solution given by Eq.~\eqref{eq012001}, 
 we obtain the solution of Eq.~\eqref{eq011903} as 
 \begin{alignat}{1}
  p_i(\phi) &= p_i(0)\e^{\nu_i \phi} \notag \\
  &+ \int_0^\phi \e^{\nu_i(\phi-\phi')}\left[
  c'_{1,i} \d w_{1,\phi'} + c'_{2,i} \d w_{2,\phi'}
  \right],\notag \\
   & \qquad \qquad (i=1,2,3,4).
 \end{alignat}

 If the real part of $\nu_i$ is non-negative,
 $\displaystyle \mathbb{E}\left[ \lim_{\phi \to +\infty} p_i (\phi)^2\right]$ diverges 
 and, consequently, heating rather than cooling occurs. 
 Every $\nu_i$ needs to have a negative real part 
 for the trapped particle to be cooled.
 The necessary and sufficient condition that every $\nu_i$ be negative is 
 \begin{equation}
  \tilde g < \frac{1}{4}\left(\tilde s + \frac{1}{\tilde s} \right).
   \label{eq012002}
 \end{equation}
 The proof is given below.
 Thus the feedback control does not work 
 under a feedback gain equal to or larger than the right-hand side of Eq.~\eqref{eq012002}.
 In the following, we assume that Eq.~\eqref{eq012002} holds 
 unless stated otherwise.

 The proof of \eqref{eq012002} goes as follows. 
 We first consider the case (i), i.e.,
 \begin{equation}
   4\tilde g - \tilde s < 0. \label{eq:032613} 
 \end{equation}
 All the $\nu_i$'s have negative real parts in this case as discussed below. 
 Let a positive number $a$ and a real number $b$ satisfy 
 \begin{equation}
  (a+\i b)^2 = {\tilde s^2 - 4 \pm 4\sqrt{4\tilde g \tilde s - \tilde s^2}}. \label{eq:032701}
 \end{equation}
 This is equivalent to 
 \begin{gather}
  a^2 - b^2 = \tilde s^2 - 4, \\
   2ab = \pm 4\sqrt{4\tilde g \tilde s - \tilde s^2}.
 \end{gather}
 Eliminating $b$ from the above two equations, we obtain 
 \begin{equation}
  a^4 - (\tilde s^2 - 4)a^2 - 4(\tilde s^2 - \tilde s \tilde g) = 0. \label{eq:032611}
 \end{equation}
 Here, we introduce $\tau = a^2 > 0$ to reduce Eq.~\eqref{eq:032611} to 
 a quadratic equation 
 \begin{equation}
   \tau^2 -(\tilde s^2 - 4)\tau - 4(\tilde s^2 - \tilde s \tilde g) = 0.\label{eq:032612}
 \end{equation}
 The left-hand side of Eq.~\eqref{eq:032612} for $\tau=0$ is negative 
 due to inequality \eqref{eq:032613}.
 Noting that the coefficient of $\tau^2$ in Eq.~\eqref{eq:032612} is positive, 
 we find that there exists only one positive solution of $\tau$.
 When $\tau=\tilde s^2$, the left-hand side of Eq.~\eqref{eq:032612} 
 is positive $4\tilde s \tilde g > 0$.
 Therefore, the solution satisfies $0<\tau < \tilde s^2$, 
 giving $a = \sqrt{\tau} < \tilde s$. 
 This ensures that 
 the real parts of $\nu_i$'s in Eqs.~\eqref{nu1}--\eqref{nu4} are negative.
 Next, we consider the case (ii), i.e., 
 \begin{equation}
   4\tilde g - \tilde s \geqslant 0. \label{eq:032614}
 \end{equation}
 In this case, only $\nu_4$ matters.
 In the same way as in the case (i), 
 we introduce a positive number $a$ and a real number $b$ that satisfy Eq.~\eqref{eq:032701}.
 Obviously, $\nu_1$ and $\nu_3$ have negative real parts.
 Since $\sqrt{4\tilde g \tilde s - \tilde s^2}$ is real in this case,
 $\tilde s^2 - 4 \pm 4\sqrt{4\tilde g \tilde s - \tilde s^2}$ is also real.
 As for $\nu_2$, $\tilde s^2 - 4 - 4\sqrt{4\tilde g \tilde s - \tilde s^2} < \tilde s^2$ always hold, 
 and hence $\nu_2$ has a negative real part.
 If and only if $\tilde s^2 - 4 + 4\sqrt{4\tilde g \tilde s - \tilde s^2} < \tilde s^2$, 
 i.e.,
 \begin{equation}
  \tilde g < \frac{1}{4}\left(\tilde s + \frac{1}{\tilde s} \right),
 \end{equation}
 $\nu_4$ can have a negative real number.
 Thus the condition \eqref{eq012002} follows.

\begin{figure}[tb]
   \centering
   \includegraphics[width=0.45\textwidth]{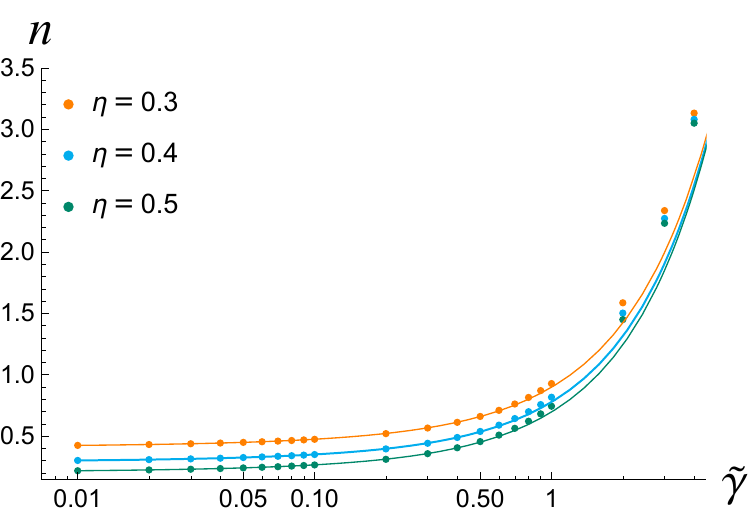}
   \caption{\label{fig:CDphoc}
  Phonon occupation number $n=\frac{E}{\hbar\omega} - \frac{1}{2}$ of the achievable energy 
  under cold damping with a band-pass filter for $\eta=0.3$ (orange), $0.4$ (blue) and $0.5$ (dark green).
  The dots are obtained from the numerical minimization of Eq.~\eqref{eq022001} with respect to $\tilde s$ and $\tilde g$
  for $\tilde \gamma = 0.01,0.02,\dots,0.09,0.1,0.2,\dots,0.9,1,2,3,4,$ 
  and the curves are obtained from Eq.~\eqref{eq:CDappendix}.
  As shown in this figure, the formula in Eq.~\eqref{eq:CDappendix} is valid for $\tilde \gamma \lesssim 1$.}
 \end{figure}

 Let us next discuss the energy of the system which is given by
 \begin{equation}
  E = \frac{P^2}{2m}+\frac{1}{2}m\omega^2 X^2 + 
  \frac{\hbar\omega}{2} \sqrt{\frac{\sqrt{1+\tilde \gamma^2}+1}{2}}. \label{eq012003}
 \end{equation}
 Here $X_0$ in Eq.~\eqref{eq011802} is set to zero and 
 the last term on the right-hand side describes the contribution 
 from the zero-point fluctuations obtained in Eq.~\eqref{eq011801}. 
 Using Eq.~\eqref{eq011904} and rewriting this energy in terms of $p_i$'s, 
 the sum of the first two terms in Eq.~\eqref{eq012003} is given by 
 \begin{equation}
  \frac{\omega^2}{2m}\sum_{i,j=1}^4 K_{ij} p_i p_j, \label{eq:032609}
 \end{equation}
 where $K_{ij}$ is a matrix defined as 
 \begin{equation}
  K_{ij} := V_{1i} V_{1j} + V_{2i} V_{2j} \quad 
  (i,j = 1,2,3,4).
 \end{equation}
 In the long-time limit, 
 the expectation value of $p_i p_j \,(i,j=1,2,3,4)$ 
 which appears in the energy expression \eqref{eq:032609} is obtained as
 \begin{equation}
  \mathbb{E}\left[ \lim_{\phi \to +\infty} p_i (\phi)p_j(\phi)\right]
  = -\frac{c'_{1i}c'_{1j} + c'_{2i} c'_{2j}}{\nu_1 + \nu_2}, \label{eq:032610}
 \end{equation}
 where we denote the average over the Wiener processes $w_1$ and $w_2$ by $\mathbb{E}$, 
 and we employ the It\^o isometry~\cite{oksendal2013stochastic} \eqref{eq012004} for $w_1$ and $w_2$.
 Substituting Eq.~\eqref{eq:032610} into Eq.~\eqref{eq:032609}, 
 we obtain an analytical expression of the energy 
 which is given by 
 \begin{widetext}
 \begin{alignat}{1}
  \notag \frac{\hbar \omega }{32 \tilde{\gamma} \tilde g \eta \tilde R^2 \tilde{s} \left(4\tilde g \tilde{s}-\tilde{s}^2-1\right)} 
  \notag \bigg\{ 
      &16\tilde g^2 \tilde R^2 \tilde{s} \left(2\tilde g \tilde{s}-\tilde{s}^2-1\right)-\tilde{\gamma}  \eta  
  \Big( -8\tilde g^2 \tilde{s} \left(\tilde{\gamma}  \left(\tilde D^2-4 \tilde D \tilde{s}-\tilde{s}^2+2\right)+2 \tilde D \tilde R (\tilde{s}^2-1)-8 \tilde R \tilde{s}\right) \\
  \notag & -2 \tilde{\gamma} \tilde g \left[\tilde D^2 \left(\tilde{s}^4+5 \tilde{s}^2-2\right)+4 \tilde D \left(\tilde{s}^3+\tilde{s}\right)+3 \tilde{s}^4+7 \tilde{s}^2-2\right]+\tilde{\gamma}  \left(\tilde D^2+1\right) \tilde{s} \left(\tilde{s}^2+1\right) \left(\tilde{s}^2+4\right) \\
    &+8\tilde g \tilde R \tilde{s} \left(\tilde{s}^2+1\right) (\tilde D \tilde{s}-2) 
  \Big) 
  \bigg\}+\frac{\hbar\omega}{2} \sqrt{\frac{\sqrt{1+\tilde \gamma^2}+1}{2}}. \label{eq022001}
 \end{alignat}
 \end{widetext}
 Minimizing the energy with respect to controllable parameters $\tilde s$ and $\tilde g$, 
 we obtain the minimal energy for given $\tilde \gamma$. 
 Dots in Fig.~\ref{fig:CDphoc} show the results of the numerical minimization of Eq.~\eqref{eq022001} 
 for $\eta=0.3,0.4$ and $0.5$.

 We assume that $\tilde g$ and $\tilde s$ depend on $\tilde \gamma$ 
 as $\tilde g \propto \tilde \gamma^\iota\,(\iota \geqslant 0)$ 
 and $\tilde s \propto \tilde \gamma^\kappa\, (\kappa\geqslant 0)$ 
 for small $\tilde \gamma$, and 
 expand the energy \eqref{eq022001} in powers of $\tilde \gamma$ 
 around $(\tilde \gamma, \tilde g, \tilde s) = (0,0,0)$.
 The leading terms are given by
 \begin{equation}
  \frac{\tilde g}{2\eta \tilde \gamma} + \frac{\tilde \gamma}{8\tilde g} 
  + \frac{\tilde \gamma}{8\tilde s} .\label{eq:020301}
 \end{equation}
 The necessary condition for each term in Eq.~\eqref{eq:020301} to be finite 
 in the limit of $\tilde \gamma \to 0$ is $\iota = 1$ and $\kappa \leqslant 1$.
 Substituting $\tilde g = \Gamma \tilde \gamma\,(\Gamma >0)$
 and $\tilde s = \Sigma \tilde \gamma^\kappa\, (\Sigma > 0)$ in Eq.~\eqref{eq022001} 
 and expanding it in powers of $\tilde \gamma$, we obtain
 \begin{equation}
  E = \frac{4\Gamma^2 + \eta}{8\Gamma\eta} 
  + \frac{\Sigma^2}{32\Gamma} \tilde \gamma^{2\kappa}
  + \frac{1}{8\Sigma} \tilde \gamma^{1-\kappa} 
  + \mathcal{O}(\tilde \gamma^{4\kappa}, \tilde \gamma^{1+\kappa}).
 \end{equation}
 Maximizing the exponent of the subleading order, we obtain $\kappa=\frac{1}{3}$.
 Then, the energy is 
 \begin{equation}
  E = \frac{4\Gamma^2 + \eta}{8\Gamma\eta} 
  + \frac{4\Gamma + \Sigma^3}{32\Gamma \Sigma} \tilde \gamma^{2/3}
  + \mathcal{O}(\tilde \gamma^{4/3}).
 \end{equation}
 Minimizing the first term and then the second term with respect to $\Gamma$ and $\Sigma$,
 we finally obtain
 \begin{equation}
  E = \hbar\omega\left[\frac{1}{2\sqrt{\eta}} 
  + \frac{3 }{16\sqrt[6]{\eta}}\tilde \gamma^{2/3}
  + \frac{5 \sqrt[6]{\eta} }{16}\tilde{\gamma}^{4/3} + \mathcal{O}(\tilde{\gamma}^2)\right]. 
  \label{eq:CDappendix}
 \end{equation}
 Here $\Gamma$ and $\Sigma$, at which the energy is minimized, 
 are $\frac{\sqrt{\eta}}{2}$ and $\sqrt[6]{\eta}$, 
 respectively.
 This approximation formula of the energy is plotted in Fig.~\ref{fig:CDphoc} (curves), 
 in excellent agreement with the exact minima (dots) for $\tilde \gamma \lesssim 1$.

 In the cold damping with a band-pass filter, 
we therewith estimate the velocity of the nanoparticle.
As described in Appendix~\ref{sec:CD},
we calculate the purity for the cold damping with a band-pass filter
by using variance of the momentum adjusted 
on the basis of this estimation of the momentum.

 {\majrev{Let us finally discuss the (co)variances of the ensemble state
  in cold damping with a band-pass filter,}
 as they are used in computation of the purity in Sec.~\ref{sec:Discussion}.
 \majrev{We do not make adjustment by using the controller's information 
 because the purity is lowered by adjusting the momentum
 by subtracting $mv_{\rm est}$ therefrom, 
 the estimated momentum based on the band-pass filter.}
 The unadjusted (co)variances are the sum of those of the complex Gaussian wavefunction~\eqref{eq:covariances-complex-Gaussian-wavefnc}
 and those of the expectation values $X$ and $P$:
 \begin{alignat}{1}
  V_x^{\rm tot} &= V_x^\psi + \lim_{t\to \infty} \mathbb{E}[X(t)^2],\\
  V_p^{\rm tot} &= V_p^\psi + \lim_{t\to \infty} \mathbb{E}[P(t)^2],\\
  V_{\rm cov}^{\rm tot} &= V_{\rm cov}^\psi + \lim_{t\to \infty} \mathbb{E}[X(t) P(t)].
 \end{alignat} 
 We note that the (co)variances are calculated in the same way as in the computation of 
 the lowest achievable energy.
 }

 \section{Achievable Energy via LQG} \label{AppLQG}

 For the purpose of comparison of the cooling methods discussed 
 in Sec.~\ref{sec:Discussion}, 
 {we follow Ref.~\cite{PhysRevA.60.2700} to calculate
  the lowest energy achievable by the LQG control
 and the (co)variances of the ensemble state prepared by it.} 
 Table \ref{tab:correspondence} shows 
 the correspondence between the notations of Ref.~\cite{PhysRevA.60.2700} and ours.
 The long-time limit of the energy of the particle 
 {and the (co)variances of the ensemble state are} given by 
 \begin{gather}
   \hspace{-30pt}\lim_{t\to \infty} \left\langle \frac{\hat p^2}{2m} + \frac{1}{2}m\omega^2 \hat x^2 \right\rangle \hspace{30pt} \notag\\ =
  \frac{V_p}{2m} + \frac{1}{2}m\omega V_x 
  +\frac{\hbar\omega}{4} \left(\tilde V_x^{\rm e} + \tilde V_p^{\rm e}\right),\label{eq:041001} \\
  {V_x^{\rm tot} = V_x + V_x^{\rm e} ,\quad 
  V_p^{\rm tot} =  V_p + V_p^{\rm e},\quad 
  V_{\rm cov}^{\rm tot} =  C + C^{\rm e}},
 \end{gather}
 where 
 \begin{alignat}{1}
  V_x &= \left(\frac{\hbar}{\sqrt{2\eta}m\omega}\right) \frac{1}{\sqrt{\xi+1}} 
  = \frac{\hbar}{2m\omega} \tilde V_x, \label{eq:conditioned_state_Vx} \\ 
  V_p &= \left(\frac{\hbar m \omega}{\sqrt{2\eta}}\right) \frac{\xi}{\sqrt{\xi+1}}
  %= \frac{\hbar m\omega}{2} \tilde V_p
  , \label{eq:conditioned_state_Vp}\\
  C &= \left(\frac{\hbar}{2\sqrt{\eta}}\right)\sqrt{\frac{\xi-1}{\xi+1}}, 
  \label{eq:conditioned_state_C}
 \end{alignat}
  are given in Eqs.~(51), (61) in Ref.~\cite{PhysRevA.60.2700}
 and 
 \begin{gather}
  \tilde V_x^{\rm e} =\frac{2m\omega}{\hbar} V_x^{\rm e} 
  = \frac{\tilde V_x^2}{r}, \quad 
  \tilde V_p^{\rm e} = \frac{2}{\hbar m \omega} V_p^{\rm e} 
  = \frac{\tilde V_x^2}{r}, \\
  \tilde C^{\rm e} = \frac{2}{\hbar} C^{\rm e} = -\frac{\tilde V_x^2}{r}, 
   \label{eq:041002}
 \end{gather} 
 are given 
 by Eqs.~(66), (67) in Ref.~\cite{PhysRevA.60.2700}.
 In Eqs.~\eqref{eq:041002}, the high-gain limit ($\mathcal{Q} \to 0$) is taken.
 Using Table \ref{tab:correspondence} and rewriting Eq.~\eqref{eq:041001} in our notations, 
 we obtain 
 \begin{equation}
  E = \hbar \omega
  \left[\frac{1}{2\sqrt{\eta}} \sqrt{\frac{\sqrt{1+\eta \tilde \gamma^2} + 1}{2}}
  + \frac{\tilde \gamma}{2(\sqrt{1+\eta \tilde \gamma^2}+1)}
  \right]. \label{appeq:LQG}
 \end{equation}

 \begin{table}[h]
   \begin{ruledtabular}
   \begin{tabular}{cccc}
  Notation of Ref.~\cite{PhysRevA.60.2700} & $k$ & $r=\dfrac{m\omega^2}{2\hbar \eta k}$ 
     & $\xi = \sqrt{1+ \dfrac{4}{\eta r^2}}$\\
  Our notation & $\dfrac{\gamma}{4}$ & $\dfrac{2}{\eta \tilde \gamma}$ & $\sqrt{1+\eta \tilde \gamma^2}$
   \end{tabular}
   \end{ruledtabular}
   \caption{\label{tab:correspondence} 
  Correspondence between notations of Ref.~\cite{PhysRevA.60.2700} and ours.}
 \end{table}

{
\section{Achievable Minimum Energy by Cold Damping with Delayed Feedback}\label{sec:CD-DaleyedFB}

{
Cold damping is most simply implemented by delayed feedback. 
This version of cold damping is widely employed in 
practice~\cite{PhysRevLett.96.043003,tebbenjohanns2021quantum,Kamba:22}.
Although the measurement outcome is filtered in
Refs.~\cite{PhysRevLett.96.043003,tebbenjohanns2021quantum,Kamba:22},
filtrations used there are intended rather to eliminate irrelevant signals 
than to extract the signal with the desired frequency. 
Thus Refs.~\cite{PhysRevLett.96.043003,tebbenjohanns2021quantum,Kamba:22} 
can be modeled by cold damping with delayed feedback discussed in the following.
}

For the conditioned state,
the long-time limits of variances of the position and the momentum, and their covariance
are given by $V_x$ and $V_p$, and $C$ in 
Eqs.~\eqref{eq:conditioned_state_Vx}--\eqref{eq:conditioned_state_C}, 
respectively \cite{PhysRevA.60.2700}. 
We note that the feedback operation does not affect these limits 
in the present case where the feedback potential is linear.
The expectation values for the conditioned state, 
denoted by 
\begin{equation}
  X_{\rm c} := \Tr[\hat \rho_{\rm c} \hat x],\quad 
  P_{\rm c} := \Tr[\hat \rho_{\rm c} \hat p], 
\end{equation}
obey the following stochastic differential equations \cite{PhysRevA.60.2700}
\begin{alignat}{1}
  \d X_{\rm c} &= \frac{P_{\rm c}}{m}\d t + \sqrt{2\eta \gamma} V_x \d W, \\
  \d P_{\rm c} &= -m\omega^2 X_{\rm c} \d t + F_{\rm fb} \d t + \sqrt{2\eta \gamma} C \d W,
\end{alignat}
where $\d W$ is a Wiener increment and 
$F_{\rm fb}$ represents feedback force.
Here the feedback force is determined on the basis of the measurement outcome 
described by \cite{PhysRevA.60.2700}
\begin{equation}
  I_{\rm c} (t) = X_{\rm c} (t) + \frac{1}{\sqrt{2 \eta \gamma}} \frac{\d W(t)}{\d t}.
\end{equation}

The cold damping with delayed feedback utilizes the following property of the harmonic oscillation: 
the velocity of the particle at time $t$ 
is proportional to the negative of the position at time $\frac{T}{4}$ earlier than $t$, 
where $T := \frac{2\pi}{\omega}$ denotes the period of the oscillation. 
We therefore adopt the feedback protocol as 
\begin{equation}
  F_{\rm fb} (t) = \tilde g m \omega^2 I\left(t - \frac{T}{4}\right),
\end{equation}
where $\tilde g$ represents the feedback gain to be optimized.

We numerically simulate the dynamics under this cold damping with delayed feedback 
to optimize the feedback gain up to two significant digits and 
to obtain the energy and the purity achievable by it.
The minimum energy and the highest purity achievable by cold damping with delayed feedback
 is plotted in Figs.~\ref{fig:compar}, \ref{fig:high_eff}, and \ref{fig:purity}, 
 {where the (co)variances for the ensemble state used to 
 calculate the purity are given by 
 \begin{alignat}{1}
  V_x^{\rm tot} &= V_x + \lim_{t\to \infty} \mathbb{E}[X_{\rm c}(t)^2] ,\\ 
  V_p^{\rm tot} &=  V_p + \lim_{t\to \infty} \mathbb{E}[P_{\rm c}(t)^2],\\
  V_{\rm cov}^{\rm tot} &=  C + \lim_{t\to \infty} \mathbb{E}[X_{\rm c}(t) P_{\rm c} (t)].
 \end{alignat} }
The numerical analysis was conducted 
according to the explicit order 2.0 weak scheme~\cite{kloeden2013numerical}.
\majrev{We compute the purity from the (co)variances 
that are not adjusted by using the controller's information 
for the following reason.
In the cold damping with delayed feedback, 
the measurement outcome~\eqref{eq:measurement-outcome}
is directly fed back to the nanoparticle.
If we adjusted the momentum with respect to the measurement outcome, 
which includes the white noise, the purity would be lower.}

}

\bibliography{LPF}

\end{document}